\documentclass[11pt,journal,onecolumn]{IEEEtran}
\usepackage[tbtags]{amsmath}
\usepackage{amsthm,amsfonts,amssymb} % AMS packages
\usepackage{bm} % bold math fonts
\usepackage[subnum]{cases} % number the formula in cases
\usepackage{bbm}

\usepackage[normalem]{ulem}%underlining for emphasis

\usepackage{graphicx}
\usepackage{subfigure}
\usepackage[svgnames]{xcolor}
\usepackage{tikz}
\usetikzlibrary{arrows}

\usepackage{multirow}%multiple rows in table
\usepackage{colortbl}%color in tables

\usepackage{cite}
\usepackage[bookmarks=false,colorlinks=true,urlcolor=cyan,citecolor=Blue,linkcolor=Green,anchorcolor=orange]{hyperref}%hyperlink

\newcommand{\Csiszar}{Csisz\'{a}r}

\newcommand{\ie}{i.e.}

\newcommand{\eg}{e.g.}

\newcommand{\iid}{{i.i.d.}\ }

\newcommand{\derivative}[2]{\frac{d\,{#1}}{d\,{#2}}}

\newcommand{\dpartial}[2]{\frac{\partial\,{#1}}{\partial\,{#2}}}

\newcommand{\cardinality}[1]{\left|#1\right|}
\newcommand{\abs}[1]{\left|#1\right|}

\newcommand{\toinf}{\rightarrow \infty}

\newcommand{\defeq}{\triangleq}
\newcommand{\setst}{:}
\newcommand{\SetDef}[2]{\left\{#1 \setst #2 \right\}}
\newcommand{\posfunc}[1]{\left\vert{#1}\right\vert^+}
\newcommand{\hvect}[2][n]{{#2}_1, {#2}_2, \ldots, {#2}_{#1}}
\newcommand{\hints}[1][n]{1, 2, \ldots, {#1}}

\newcommand{\nnreals}{\mathbb{R}_+}
\newcommand{\reals}{\mathbb{R}}

\newcommand{\pints}{\mathbb{Z^+}}
\newcommand{\nnints}{\mathbb{Z_+}}

\newcommand{\eps}{\varepsilon}

\newcommand{\indicator}[1]{\mathbbm{1}_{#1}}

\newcommand{\floor}[1]{\lfloor{#1}\rfloor}

\newcommand{\onenorm}[1]{\|#1\|_1}

\newcommand{\infnorm}[1]{\|#1\|_{\infty}}

\newcommand{\avelog}[1]{\dfrac{1}{#1}\log}

\newcommand{\navelog}{\avelog{n}}

\newcommand{\BigO}[1]{O\left({#1}\right)}
\newcommand{\BigOmega}[1]{\Omega\left({#1}\right)}
\newcommand{\BigTheta}[1]{\Theta\left({#1}\right)}
\newcommand{\E}[1]{\mathbb{E}\left[{#1}\right]}
\newcommand{\Ep}[2]{\mathbb{E}_{#1}\left[{#2}\right]}
\newcommand{\Var}[1]{\mathrm{Var}\left[{#1}\right]}

\newcommand{\Qinv}[1]{Q^{-1}\left({#1}\right)}

\newcommand{\polyn}{\mathrm{poly}(n)}

\newcommand{\Set}[1]{\left\{{#1}\right\}}

\newcommand{\chdef}[3]{#1: {#2}\rightarrow{#3}}
\newcommand{\chio}[3]{#1\left(\left.{#2}\,\right\vert{#3}\right)}
\newcommand{\chin}[2]{\chio{#1}{\cdot}{#2}}

\newcommand{\DMC}[1][W]{DMC $(\cX, \cY, {#1})$}

\newcommand{\chion}[4][n]{{#2}^{#1}\left(\left.#3\right|#4\right)}
\newcommand{\chinn}[3][n]{{#2}^{#1}\left(\cdot|#3\right)}

\newcommand{\Entropy}[1]{H\left(#1\right)}
\newcommand{\CondEntropy}[2]{H\left(#1|#2\right)}

\newcommand{\KLD}[2]{D\left({#1}\,\middle\|\,#2\right)}
\newcommand{\CKLD}[3]{D\left({#1}\,\middle\|\,#2|#3\right)}

\newcommand{\MIXY}[2]{I\left({#1};{#2}\right)}
\newcommand{\MIPW}[2]{I\left({#1},{#2}\right)}

\newcommand{\MIxy}[2]{I\left({#1}\wedge{#2}\right)}

\newcommand{\InfoVar}[2]{U\left({#1},{#2}\right)}
\newcommand{\CondInfoVar}[2]{V\left({#1},{#2}\right)}
\newcommand{\Vmax}{V_\mathrm{max}}
\newcommand{\Vmin}{V_\mathrm{min}}

\newcommand{\KLDV}[2]{V\left(\left.{#1}\,\right\|#2\right)}
\newcommand{\CKLDV}[3]{V\left(\left.{#1}\,\right\|#2|#3\right)}

\newcommand{\PDSpace}[1]{\cP\left(#1\right)}
\newcommand{\PDSpaceX}{\PDSpace{\cX}}

\newcommand{\PDSpaceN}[1]{\cP_n\left(#1\right)}
\newcommand{\PDSpaceXN}{\PDSpaceN{\cX}}

\newcommand{\PDProd}[2]{{#1}\cdot{#2}} % product of distributions

\newcommand{\TypeSetn}[2][n]{\mathcal{T}^{#1}_{#2}}

\newcommand{\TypeShell}[2]{\mathcal{T}_{#1}\left(#2\right)}
\newcommand{\TypeShelln}[3][n]{\mathcal{T}^{#1}_{#2}\left(#3\right)}

\newcommand{\Pe}{P_\mathrm{e}}

\newcommand{\Pef}[1]{\Pe\left(#1\right)}

\newcommand{\Prob}[1]{\mathbb{P}\left[{#1}\right]}

\newcommand{\nn}{\nonumber}

\DeclareMathAlphabet{\mathbsf}{OT1}{cmss}{bx}{n}% bold sans serif
\DeclareMathAlphabet{\mathssf}{OT1}{cmss}{m}{sl}% slanted sans serif

\DeclareSymbolFont{bsfletters}{OT1}{cmss}{bx}{n}  
\DeclareSymbolFont{ssfletters}{OT1}{cmss}{m}{n}
\DeclareMathSymbol{\bsfGamma}{0}{bsfletters}{'000}
\DeclareMathSymbol{\ssfGamma}{0}{ssfletters}{'000}
\DeclareMathSymbol{\bsfDelta}{0}{bsfletters}{'001}
\DeclareMathSymbol{\ssfDelta}{0}{ssfletters}{'001}
\DeclareMathSymbol{\bsfTheta}{0}{bsfletters}{'002}
\DeclareMathSymbol{\ssfTheta}{0}{ssfletters}{'002}
\DeclareMathSymbol{\bsfLambda}{0}{bsfletters}{'003}
\DeclareMathSymbol{\ssfLambda}{0}{ssfletters}{'003}
\DeclareMathSymbol{\bsfXi}{0}{bsfletters}{'004}
\DeclareMathSymbol{\ssfXi}{0}{ssfletters}{'004}
\DeclareMathSymbol{\bsfPi}{0}{bsfletters}{'005}
\DeclareMathSymbol{\ssfPi}{0}{ssfletters}{'005}
\DeclareMathSymbol{\bsfSigma}{0}{bsfletters}{'006}
\DeclareMathSymbol{\ssfSigma}{0}{ssfletters}{'006}
\DeclareMathSymbol{\bsfUpsilon}{0}{bsfletters}{'007}
\DeclareMathSymbol{\ssfUpsilon}{0}{ssfletters}{'007}
\DeclareMathSymbol{\bsfPhi}{0}{bsfletters}{'010}
\DeclareMathSymbol{\ssfPhi}{0}{ssfletters}{'010}
\DeclareMathSymbol{\bsfPsi}{0}{bsfletters}{'011}
\DeclareMathSymbol{\ssfPsi}{0}{ssfletters}{'011}
\DeclareMathSymbol{\bsfOmega}{0}{bsfletters}{'012}
\DeclareMathSymbol{\ssfOmega}{0}{ssfletters}{'012}
\newcommand{\cA}{\mathcal{A}}

\newcommand{\cB}{\mathcal{B}}
\newcommand{\hB}{\hat{B}}

\newcommand{\cC}{\mathcal{C}}

\newcommand{\cE}{\mathcal{E}}

\newcommand{\cJ}{\mathcal{J}}

\newcommand{\cM}{\mathcal{M}}

\newcommand{\cP}{\mathcal{P}}

\newcommand{\bS}{\mathbf{S}}
\newcommand{\cS}{\mathcal{S}}
\newcommand{\hS}{\hat{S}}

\newcommand{\brV}{\bar{V}}
\newcommand{\bX}{\mathbf{X}}
\newcommand{\cX}{\mathcal{X}}

\newcommand{\bY}{\mathbf{Y}}
\newcommand{\cY}{\mathcal{Y}}

\newcommand{\tlZ}{\tilde{Z}}

\newcommand{\hmm}{\hat{m}}

\newcommand{\bx}{\mathbf{x}}

\newcommand{\by}{\mathbf{y}}

\newcommand{\tlLambda}{\tilde{\Lambda}}

\definecolor{shadecolor}{rgb}{0.90,0.90,0.90}
\definecolor{MyGray}{rgb}{0.96,0.97,0.98}
\definecolor{boxcolor}{rgb}{0.90,0.90,0.90}
\definecolor{darkblue}{RGB}{0,0,128}
\definecolor{darkred}{RGB}{128,0,0}
\definecolor{darkgreen}{RGB}{0,128,0}
\definecolor{mblue}{RGB}{0,0,204}
\definecolor{mred}{RGB}{204,0,0}

\usepackage{aliascnt}
\usepackage[capitalize]{cleveref}% must be the last usepackage

\creflabelformat{equation}{\hspace{-0.5ex}(#2{}#1{}#3)}
\crefname{equation}{}{}
\Crefname{equation}{}{}
\crefname{thm}{theorem}{theorems}
\Crefname{thm}{Theorem}{Theorems}
\crefname{prop}{proposition}{propositions}
\Crefname{prop}{Proposition}{Propositions}
\crefname{figure}{fig.}{figures}
\Crefname{figure}{Fig.}{Figures}
\crefname{defn}{definition}{definitions}
\Crefname{defn}{Definition}{Definitions}
\crefname{fact}{fact}{facts}
\Crefname{fact}{Fact}{Facts}
\crefname{app}{appendix}{appendices}
\Crefname{app}{Appendix}{Appendices}

\usepackage{tikz}
\usepackage{pgfplots}
\usepackage{nicefrac}

\newcommand{\ED}{\mathcal{E}\!(D)}

\newcommand{\Types}{\mathcal{T}}

\newcommand{\flognn}{\frac{\log n}{n}}
\newcommand{\Oflognn}{\BigO{\frac{\log n}{{n}}}}

\newcommand{\Oflognrn}{\BigO{\frac{\log n}{\sqrt{n}}}}

\newcommand{\comma}{,}

\usetikzlibrary{patterns}
\creflabelformat{equation}{\hspace{-0.5ex}(#2#1#3)}

\renewcommand{\PDProd}[2]{[{#1}\times{#2}]} % product of distributions

\newcommand{\rem}[1]{}

\newtheorem{theorem}{Theorem}

\newaliascnt{lemma}{theorem}
\newtheorem{lemma}[lemma]{Lemma}
\aliascntresetthe{lemma}

\newaliascnt{prop}{theorem}

\aliascntresetthe{prop}

\newaliascnt{corollary}{theorem}
\newtheorem{corollary}[corollary]{Corollary}
\aliascntresetthe{corollary}

\newaliascnt{remark}{definition}
\newtheorem{remark}[remark]{Remark}
\aliascntresetthe{remark}

\newcommand{\bs}{\mathbf{s}}

\newcommand{\Pch}{\Phi}
\newcommand{\Pchn}[1][n]{\Phi_{#1}}
\renewcommand{\PDProd}[2]{[{#1}\times{#2}]} % product of distributions

\begin{document}
\pgfplotsset{every axis/.append style={
    axis x line=bottom,
    axis y line=left,
    tick label style={font=\large},
    label style={font=\large},
    ticks=both,
}}

\title{\huge The Dispersion of Joint Source-Channel Coding} %Excess Distortion Dispersion }

\author{Da~Wang, %
 Amir~Ingber, %~\IEEEmembership{Member,~IEEE,}%
 Yuval~Kochman%,~\IEEEmembership{Member,~IEEE}%
}

%\author{
%\authorblockN{Da Wang}
%\authorblockA{EECS Dept.,
%MIT\\
%Cambridge, MA 02139, USA \\
%Email: dawang@mit.edu}
%\and
%\authorblockN{Amir Ingber}
%\authorblockA{Dept. of EE-Systems,
%TAU\\
%Tel Aviv 69978, Israel\\
%Email: ingber@eng.tau.ac.il}
%\and
%\authorblockN{Yuval Kochman}
%\authorblockA{EECS Dept.,
%MIT\\
%Cambridge, MA 02139, USA \\
%Email: yuvalko@mit.edu}
%}

\maketitle

\begin{abstract}
In this work we investigate the behavior of the distortion threshold that can be guaranteed in joint source-channel coding, to within a prescribed excess-distortion probability. We show that the gap between this threshold and the optimal average distortion is governed by a constant that we call the joint source-channel dispersion. This constant can be easily computed, since it is the sum of the source and channel dispersions, previously derived. The resulting performance is shown to be better than that of any separation-based scheme. For the proof, we use unequal error protection channel coding, thus we also evaluate the dispersion of that setting.

\end{abstract}

%\newpage
%\TODO{DW: this page is for our reference only and will be removed later.}
%\tableofcontents
%\newpage

\section{Introduction}
\label{sec:intro}
One of the most basic results of Information Theory, joint source-channel coding, due to
Shannon~\cite{ShannonRDF}, states that
in the limit of large block-length $n$, a discrete memoryless source with distribution $P$ can
be sent through a discrete memoryless channel with transition distribution $W$ and reconstructed
with some expected average distortion $D$, as long as
\begin{align} 
    \label{eq:RDC}
	R(P,D) <  \rho C(W),
\end{align}
where $R(P,D)$ is the rate-distortion function of the source, $C(W)$ is the channel capacity and
the bandwidth expansion ratio $\rho$ is the number of channel uses per source sample. We denote by $D^*=D^*(P,W,\rho)$ the
distortion satisfying \Cref{eq:RDC} with equality, known as the optimal performance theoretically
attainable (OPTA).  Beyond the \emph{expected} distortion, one may be interested in ensuring that
the distortion for one source block is below some threshold. To that end, we see an \emph{excess
distortion} event $\ED$ as 
\begin{equation}
    \label{eq:ED}
    \ED 
    \defeq \{ d(\bS,\hat\bS)>D \},
\end{equation}
where 
\begin{align} 
    \label{eq:distortion} 
    d(\bs,\hat\bs) \defeq \frac{1}{n}\sum_{i=1}^n d(s_i,\hat s_i)
\end{align} 
is the distortion between the source and reproduction words $\bs$ and $\hat \bs$.

We are interested in the probability of this event as a function of the block length. We note that
two different approaches can be taken. In the first, the distortion threshold is fixed to some
$D\geq D^*$ and one considers how the excess-distortion probability $\eps$ approaches zero as the
block length $n$ grows. This leads to the joint source-channel excess-distortion exponent:
\cite{CsiszarLosslessJointExponent,CsiszarJointExponent}
\begin{align} 
    \label{eq:JSCC_exponent}
    \eps(n) \cong \exp\{-n \cdot  E(P,W,\rho,D) \}.
\end{align}

One may ask an alternative question: for given excess distortion probability $\eps$, let $D_n$ be the optimal
(minimal) distortion threshold that can be achieved at blocklength $n$. How does the sequence $D_n$ approach $D^*$? In this work we show, that the sequence behaves as:
\begin{align}
    \label{eq:JSCC_dispersion}
    R(P,D_n) \cong \rho C(W) -  \sqrt \frac{V_J(P,W,\rho)}{n} Q^{-1} (\eps),  
\end{align} where $Q^{-1}(\cdot)$ is the inverse of the Gaussian cdf. We coin $V_J(P,W,\rho)$ the
joint source-channel coding (JSCC) dispersion.

Similar problems have been stated and solved in the context of channel coding and lossless source
coding in\cite{Strassen62_Asymptotische}. In \cite{PolyanskiyPVFiniteLength10} the channel
dispersion result is tightened and extended, while in \cite{AmirYuvalDCC} (see also
\cite{Kostina_Source_ISIT}) the parallel lossy source coding result is derived. In source coding,
the rate redundancy above the rate-distortion function (or entropy in the lossless case) is
measured, for a given excess-distortion probability $\eps$:
\begin{align}
    \label{eq:S_dispersion}
    R_n \cong R(P,D) +  \sqrt \frac{V_S(P,D)}{n} Q^{-1} (\eps), 
\end{align}
where $V_S(P,D)$ is the source-coding dispersion. In channel coding, it is the rate gap below capacity, for a given error probability $\eps$:
\begin{align}
    \label{eq:C_dispersion}
    R_n \cong C(W) -  \sqrt \frac{V_C(W)}{n} Q^{-1} (\eps),  
\end{align}
where $V_C(W)$ is the channel-coding dispersion. We show that the JSCC dispersion is related to the
source and channel dispersions by the following simple formula (subject to certain regularity conditions):
\begin{align}
    \label{eq:dispersion_formula}
    V_J(P,W,\rho) = V_S(P,D^*) + \rho \cdot V_C(W).
\end{align}

The achievability proof 
of \Cref{eq:dispersion_formula}
is closely related to that of \Csiszar\ for the exponent
\cite{CsiszarJointExponent}. Namely, multiple source codebooks are mapped into an unequal error
protection channel coding scheme. The converse proof combines the strong channel coding
converse~\cite{DueckKorner} with the $D$-covering of a type class (\eg,~\cite{ZhangYangWei97}).

The rest of the paper is organized as follows. \Cref{sec:notations} defines the notations.
\Cref{sec:UEP} revisits the channel coding problem, and extend the dispersion result
\Cref{eq:C_dispersion} to the unequal error protection (UEP) setting. \Cref{sec:JSCC} uses this
framework to prove our main JSCC dispersion result. Then \Cref{sec:separation} shows the dispersion
loss of separation-based schemes. Finally in \Cref{sec:lossless} we consider a formulation where the distortion ratios are fixed but
the bandwidth expansion ratio $\rho$ varies with $n$, and apply it to the lossless JSCC dispersion
problem.

\section{Notations}
\label{sec:notations}
This paper uses lower case letters (\eg\ $x$) to denote a particular value of the
corresponding random variable denoted in capital letters (\eg\ $X$). Vectors are denoted in bold
(\eg\ $\bx$ or $\bX$).
caligraphic fonts (\eg\ $\cX$) represent a set and $\PDSpaceX$ for all the probability distributions
on the alphabet $\cX$.
We use $\nnints$ and $\nnreals$ to denote the set of non-negative integer and real numbers
respectively.

Our proofs make use of the method of types, and follow the notations in~\cite{CsiszarBook}.
Specifically, the \emph{type} of a sequence $\bx$ with length $n$ is denoted by $P_\bx$, where the type is
the empirical distribution of this sequence, \ie,
$
    P_\bx(a) = N(a|\bx)/n
    \,
    \forall a \in \cX,
    %\label{}
$
where $N(a|\bx)$ is the number of occurrences of $a$ in sequence $\bx$. The subset of the probability distributions $\PDSpaceX$ that
can be types of $n$-sequences is denoted as
\begin{equation}
    \PDSpaceXN \defeq
    \SetDef{P \in \PDSpaceX}{
        n P(x) \in \nnints,\,
        \forall x \in \cX
        }
    \label{eq:all_types}
\end{equation}
and sometimes $P_n$ is used to emphasize the fact that $P_n \in \PDSpaceXN$.
A \emph{type class} $\TypeSetn{P_\bx}$ is defined as the set of sequences that have type $P_\bx$.
Given some sequence $\bx$, a sequence $\by$ of the same length has \emph{conditional type} $P_{\by|\bx}$ if
$N(a,b|\bx,\by) = P_{\by|\bx}(a|b) N(a|\bx)$.
Furthermore, the random variable corresponding to the conditional type of
a random vector $\bY$ given $\bx$ is denoted as $P_{\bY|\bx}$.
In addiiton, the possible conditional type given an input distribution $P_\bx$ is denoted as 
\begin{align*}
    \PDSpaceN{\cY|P_\bx}
    &\defeq
    \SetDef{
        P_{\by|\bx}
    }{
        %P_{\by|\bx}(a|b) 
        %= 
        %N(a, b|\bx, \by)
        %/
        %(n P_\bx(a))
        %\text{ for all } 
        %\by \in \cY^n.
        P_\bx
        \times
        P_{\by|\bx} 
        \in
        \PDSpaceN{\cX \times \cY}
    }.
\end{align*}
    
A discrete memoryless channel (DMC) $\chdef{W}{\cX}{\cY}$ is defined with its input alphabet
${\cX}$, output alphabet ${\cY}$, and conditional
distribution $\chin{W}{x}$ of output letter $Y$ when the channel input letter $X$ equals $x\in\cX$.
Also, we abbreviate $\chin{W}{x}$ as $W_x(\cdot)$ for notational simplicity. 
We define mutual information as
\[
\MIPW{\Pch}{W} = \sum_{x,y} \Pch(x) W(y|x) \log \frac{ \Pch(x) W(y|x) }{ \Pch W(y) },
\]
and the channel capacity is given by
\[
C(W) = \max_{\Pch} \MIPW{\Pch}{W},
\]
and the set of capacity-achieving distributions is
$\Pi(W) \defeq \SetDef{\Pch}{\MIPW{\Pch}{W} = C(W)}$.

A discrete memoryless source (DMS) is defined with source alphabet $\cS$, reproduction alphabet
${\hat \cS}$, source distribution $P$ and a distortion measure $d:\cS\times\hat\cS \rightarrow
\nnreals$. Without loss of generality, we assume that for any $s\in\cS$ there is $\hat s\in\hat\cS$
such that $d(s,\hat s)=0$. The rate-distortion function (RDF) of a DMS $(\cS,\hat\cS,P,d)$ is given
by
\begin{equation*} 
    R(P,D) = \min_{\Lambda: \substack{E_{P,\Lambda} d(S,\hat S) \leq D}} I(P,\Lambda), 
\end{equation*}
where $I(P,\Lambda)$ is the mutual information over a channel with input distribution $P(S)$ and
conditional distribution $\Lambda: {\cS} \rightarrow \hat{\cS}$.

A discrete memoryless joint source-channel coding (JSCC) problem consists of a DMS
$(\cS,\hat\cS,P,d)$, a DMC $\chdef{W}{\cX}{\cY}$ and a \emph{bandwidth expansion factor} $\rho \in
\nnreals$.  
A JSCC scheme is comprised of an encoder mapping $f_{J;n}: \cS^n \rightarrow \cX^{\floor{\rho n}}$
and decoder mapping $g_{J;n}: \cY^{\floor{\rho n}} \rightarrow \hat\cS^n$.  Given a source block
$\bs$, the encoder maps it to a sequence $\bx=f_{J;n}(\bs) \in \cX^{\floor{\rho n}}$ and transmits
this sequence through the channel. The decoder receives a sequence $\by \in \cY^{\floor{\rho n}}$
distributed according to $W(\cdot|\bx)$, and maps it to a source reconstruction $\hat \bs$. The
corresponding distortion is given by \Cref{eq:distortion}.

For our analysis, we also define the following information
quantities~\cite{PolyanskiyPVFiniteLength10}: given input distribution $\Pch$ and channel $W$, we
define the information density of a channel as
\begin{equation*}
    i(x, y)
    \defeq \log \derivative{\chio{W}{y}{x}}{\Pch W(y)}
    = \derivative{\MIPW{\Pch}{W}}{W}
    = \dpartial{\MIPW{\Pch}{W}}{W},
    \label{eq:info_density}
\end{equation*}
divergence variance as
\begin{equation*}
    \KLDV{\Pch}{\Psi}
    = \sum_{x \in \cX} \Pch(x) \left[ \log\frac{\Pch(x)}{\Psi(x)} \right]^2 - [\KLD{\Pch}{\Psi}]^2,
\end{equation*}
unconditional information variance as
\begin{equation*}
    \InfoVar{\Pch}{W}
    \defeq
    \Var{ i(X, Y)}
    = \KLDV{\Pch \times W}{\Pch \times \Pch W},
    \label{eq:info_var}
\end{equation*}
where $X \times Y$ has joint distribution $\PDProd{\Pch}{W}$,
conditional information variance as
\begin{equation*} \begin{split}
    \CondInfoVar{\Pch}{W}
    &\defeq\E{\Var{ i(X, Y) | X}}
    \label{eq:cond_info_var}
    \\
    &= \CKLDV{\Pch}{\Pch W}{\Pch}
    \\
    &= \sum_{x \in \cX} \Pch(x) \Bigl\{
        \sum_{y \in \cY} \chio{W}{y}{x}
        \left[
        \log\frac{\chio{W}{y}{x}}{\Pch W(y)}
        \right]^2 \Bigr. \\
        & \quad
        - \Bigl.
        [\KLD{W_x}{\Pch W}]^2 \Bigr\} ,
\end{split} \end{equation*}
and maximal/minimal conditional information variance as
\begin{align*}
    \Vmax(W)
    &\defeq \max_{\Pch \in \Pi(W)} \CondInfoVar{\Pch}{W},
    \\
    \Vmin(W)
    &\defeq \min_{\Pch \in \Pi(W)} \CondInfoVar{\Pch}{W}.
    %\label{}
\end{align*}
For simplicity, we assume all channels in this paper satisify $\Vmin > 0$, which holds for most
channels (see~\cite[Appendix H]{PolyanskiyPVFiniteLength10} for detailed discussion).

In this paper, we use the notation $\BigO{\cdot}$, $\BigOmega{\cdot}$ and $\BigTheta{\cdot}$, where
$f(n) = \BigO{g(n)}$ if and only if
$
\limsup_{n\toinf} \abs{\frac{f(n)}{g(n)}} < \infty,
$
$f(n) = \BigOmega{g(n)}$ if and only if
$\liminf_{n\toinf} \abs{\frac{f(n)}{g(n)}} \geq 1$,
and
$f(n) = \BigTheta{g(n)}$ if and only if $f(n) = \BigO{g(n)}$ and $f(n) = \BigOmega{g(n)}$.
In addition, $f(n) \leq O(g(n))$ means that $f(n) \leq c g(n)$ for some $c>0$ and
sufficiently large $n$.
And we use the notation $\polyn$ to denote a sequence of numbers that is polynomial in $n$, \ie,
$\polyn = \BigTheta{n^d}$ if the polynomial has degree $d$.

%Note that
%\begin{align*}
%    \CondInfoVar{P}{W} + \Ep{P}{\left[ \KLD{W_x}{PW} \right]^2}
%    &=
%    \InfoVar{P}{W} + \left[ \MIPW{P}{W} \right]^2
%\end{align*}
%and
%\begin{align*}
%    \left[  \MIPW{P}{W} \right]^2
%    &= \left[ \Ep{P}{\KLD{W_x}{PW}} \right]^2
%    \\
%    &\leq \Ep{P}{ \left[\KLD{W_x}{PW}\right]^2}
%\end{align*}
%Hence

%\section{Second-Order Analysis of Functions of Distributions}
%\label{sec:Redundancy}
%\input{Redundancy}

\section{The Dispersion of UEP Channel Coding}
\label{sec:UEP}
In this section we introduce the dispersion of unequal error protection (UEP) coding. We use
this framework in the next section to prove our main JSCC result, though we directly use 
one lemma proven here instead of the UEP dispersion theorem%
\footnote{In this section we use $n$ to denote the channel code block length, while in
\Cref{sec:JSCC,sec:separation,sec:lossless} we use $m = \floor{\rho n}$ as the channel code block length in the JSCC setting.}%
.

Given $k$ classes of messages $\hvect[k]{\cM}$, where $\cardinality{\cM_i} = N_i$,
we can represent a message $m \in \cM \defeq \cup_i \cM_i$ by its \emph{class} $i$ and
\emph{content} $j$, \ie, $m = (i,j)$,
where $i \in \Set{\hints[k]}$ and $j \in \Set{\hints[N_i]}$.
A scheme is comprised of an encoding function $f_{C;n}: \cM \rightarrow \cX^n$ and a decoder mapping $g_{C;n}: \cY^n \rightarrow \cM$.
The error probability for message $m$ is
$\Pef{m} \defeq \Prob{ \hmm \neq m}$, where $\hmm$ is the decoder output.
We say that a scheme $(f_{C;n}, g_{C;n})$ is a \emph{UEP scheme} with
error probabilities
$\hvect[k]{e}$
and rates
$\hvect[k]{R}$
if
\begin{equation*}
    \Pef{m=(i,j)} \leq e_i
  %  \quad
  %  \text{for all }
  %  i \in \Set{\hints[k]},
   % j \in \Set{\hints[N_i]},
    %\label{}
\end{equation*}
for all messages, and
\begin{equation*}
    R_i = \navelog N_i
    \quad
    \text{for all }
    i \in \Set{\hints[k]},
    %\label{}
\end{equation*}
where $n$ is the block length. We denote the codewords for message set $M_i$ by $\cA_i$, \ie,
\[
\cA_i \defeq \Set{f_{C;n}(m = (i,j)), j=1,2,\cdots,N_i}.
\]

As discussed in~\cite{PolyanskiyPVFiniteLength10}, dispersion gives a meaningful characterization on
the rate loss at a certain block length and error probability. Here, we show that similar results
hold for UEP channel codes.

\begin{theorem}[UEP Dispersion, Achievability]
    \label{thm:UEP_achievable}
    Given a \DMC, a sequence of integers $k_n = \polyn$,
    an infinite sequence of real numbers $\Set{\eps_i \in (0,1), i \in \pints}$
    and
    an infinite sequence of (not necessarily distinct) distributions $\Set{ \Pch^{(i)} \in \PDSpaceX, i \in \pints}$,
    if $\CondInfoVar{\Pch^{(i)}}{W} > 0\; \forall\; i$ , then
  %  for $n > n_0(\cardinality{\cX}, \cardinality{\cY}, \left\{ k_n \right\})$,
    there exists a sequence of UEP schemes with $k_n$ classes of messages and
    error probabilities $e_i \leq \eps_i$ such that for all $1\leq i\leq k_n$,
    \begin{equation}
        \label{eq:UEP_achievable_rate}
        R_i
        = \MIPW{\Pch^{(i)}}{W} - \sqrt{\frac{V_i}{n}}Q^{-1}(\eps_i) + \BigO{\frac{\log n}{n}},
    \end{equation}
    where $V_i \defeq \CondInfoVar{\Pch^{(i)}}{W}$ is the conditional information variance in 
    \Cref{eq:cond_info_var}.
\end{theorem}

The following corollary is immediate, substituting types $\Set{\Phi_i \in \Pi(W)}$.

\begin{corollary}
    \label{coro:UEP_at_C}
    In the setting of \Cref{thm:UEP_achievable}, there exists a sequence of UEP codes with error probabilities $e_i \leq \eps_i$ such that
    \begin{equation*}
        R_i
        = C(W) - \sqrt{\frac{V_{C_i}}{n}}\Qinv{\eps_i} + \BigO{\frac{\log n}{n}},
        %\quad
        %i = 1, 2, \ldots, k_n
        %\label{}
    \end{equation*}
    where 
    \begin{equation*}
        V_{C_i} =
        \begin{cases}
        \Vmin(W)
        & \eps_i \leq \frac{1}{2}
        \\
        \Vmax(W)
        & \eps_i > \frac{1}{2}
        \end{cases}.
        \label{eq:V_C_i}
    \end{equation*}
\end{corollary}

\begin{remark} \label{rem:non_uniform} 
    In the theorem, the coefficient of the correction term $\BigO{\log n / n}$ is unbounded for
    error probabilities that approach zero or one. 
\end{remark}

\begin{remark} 
    \label{rem:cumulative} 
    In the theorem, the message classes are cumulative, i.e., for each codeword length $n$, $k_n$
    message classes are used, which include the $k_{n-1}$ classes used for $n-1$. Trivially, at
    least the same performance is achievable where only the message classes $k_{n-1}+1,\ldots,k_n$
    are used. Thus, the theorem also applies to disjoint message sets, as long as their size is
    polynomial in $n$.
\end{remark}

\begin{remark}
    The rates of \Cref{coro:UEP_at_C} are also necessary (up to the correction term).
    That is, any UEP code with error probabilities $\hvect[k_n]{e}$ such that
    $e_i \leq \eps_i $ must satisfy
    \begin{equation*}
        R_i
        \leq C(W) - \sqrt{\frac{V_{C_i}}{n}}\Qinv{\eps_i} + \BigO{\frac{\log n}{n}}.
    \end{equation*}
    This is straightforward to see, as Theorem 48
    of~\cite{PolyanskiyPVFiniteLength10} shows that this is a bound in the single-codebook case.
\end{remark}

\begin{remark}
    When taking a single codebook, i.e. $k_n=1$ for all $n$, \Cref{coro:UEP_at_C} reduces to the
    achievability part of the channel dispersion result~\cite[Theorem 49]{PolyanskiyPVFiniteLength10}.
    However, we have taken a slightly different path: we use constant-composition codebooks,
    resulting in the conditional information variance $\CondInfoVar{\Pch}{W}$, rather than \iid
    codebooks which result in the generally higher (worse) unconditional information variance.  As
    discussed in \cite{PolyanskiyPVFiniteLength10}, these quantities are equal when a
    capacity-achieving distribution is used, but a scheme achieving $\CondInfoVar{\Pch}{W}$ may have
    an advantage under a cost constraint. Furthermore, we feel that our approach is more insightful,
    since it demonstrates that the stochastic effect that governs the dispersion is in the channel
    realization only, and not in the channel input (dual to the source dispersion being set by the
    source type only). 
\end{remark}

The proof of \Cref{thm:UEP_achievable} is based on the same construction used for the UEP exponent
in~\cite{CsiszarLosslessJointExponent}. A decoder that operates based on empirical mutual
information (with varying threshold according to the codebook) is used, and if there is a unique
codeword that has high enough empirical mutual information, it is declared; otherwise an error will
be reported. This decoding rule may introduce two types of errors: the empirical mutual information
for the actual codeword is not high enough, or the empirical mutual information for a wrong codeword
is too high.

The following two lemmas address the effect of these error events.  \Cref{lemma:rate_redundancy}
shows that the empirical mutual information of the correct codeword is approximately normal
distributed via the Central Limit Theorem, hence the probability of the first type of error (the
empirical mutual information falls below the expected mutual information) is governed by the
$Q$-function, from which we can obtain expression for the rate redundancy w.r.t. empirical mutual
information.  \Cref{lemma:UEP_error2} shows that if we choose the codebook properly, the probability
of the second type of error can be made negligible, relative to the probability of the first type of
error.

\begin{lemma}[Rate redundancy]
    \label{lemma:rate_redundancy}
    For a \DMC, given a an arbitrary distribution $\Pch \in \PDSpaceX$ with $\CondInfoVar{\Pch}{V} >
    0$, and a fixed probability $\eps$,
    let $\Pchn \in \PDSpaceXN$ be an $n$-type that approximates $\Pch$ as 
    \begin{equation}
        \|\Pch - \Pchn\|_\infty\leq \frac{1}{n}.
        \label{eq:type_proximity}
    \end{equation}
    Let the \emph{rate redundancy} $\Delta R$ be the infimal value such that for $\bx \in
    \TypeSetn{\Pchn}$,
    \begin{equation}
        \label{eq:rate_redundancy_eps}
        \Prob{
            \MIPW{\Pchn}{P_{\bY|\bx}} \leq \MIPW{\Pch}{W} - \Delta R, \bY \sim \chinn{W}{\bx} 
        }
        = \eps,
    \end{equation}
    then
    \begin{equation}
        \Delta R
        =
        \sqrt{\frac{V(\Pch,W)}{n}} \Qinv{\eps}
        + \BigO{\frac{\log n}{n}}.
        \label{eq:rate_redundancy}
    \end{equation}
    Furthermore, the result holds if we replace \Cref{eq:rate_redundancy_eps} with
    \begin{equation}
        \label{eq:rate_redundancy_eps2}
        \Prob{
            \MIPW{\Pchn}{P_{\bY|\bx}} \leq \MIPW{\Pch}{W} - \Delta R, \bY \sim \chinn{W}{\bx} 
        }
        = \eps + \delta_n,
    \end{equation}
    as long as 
    $\delta_n = \BigO{\frac{\log n}{\sqrt{n}}}$.
\end{lemma}
\begin{IEEEproof}[Proof sketch for \Cref{lemma:rate_redundancy}]
Applying Taylor expansion to the empirical mutual information $I(\Pchn,P_{\bY|\bx})$, where
$\bY$ is the channel output corresponding to channel input $\bx$, we have
\begin{align*}
    I(\Pchn,P_{\bY|\bx})
    &\approx I(\Pchn,W) \\
    &+ \sum_{x\in\cX,y\in\cY} (P_{\bY|\bx}(y|x) - W(y|x)) I'_W(y|x),
\end{align*}
where the higher order terms only contribute to the correction term in the desired result,
and \[I'_W(y|x) \triangleq\left.\frac{\partial I(\Pchn,V)}{\partial V(y|x)}\right|_{V = W}. \]
These first order terms can be represetned by sum of independent random variables with total variance
${V(\Pchn,W)}/{n}$ and finite third moment, which faciliates the application of Berry-Esseen theorem
(see, e.g.,  \cite[Ch. XVI.5]{Feller1971}) and gives
\begin{align*}
    &\,\,\Prob{ \MIPW{\Pchn}{P_{\bY|\bx}} \leq \MIPW{\Pchn}{W} - \Delta R }
    \nn
    \\
    \approx&
    \,\,
    Q\left( \left( \Delta_n + \Delta R \right) \sqrt{\frac{n}{{V}}} \right),
\end{align*}
where $\Delta_n = \BigO{\log n/n}$.
Finally, we can show that given \Cref{eq:type_proximity},
$\abs{{V(\Pch,W)} -{V(\Pchn,W)}}$
and\\
$\abs{{\MIPW{\Pch}{W}} -{\MIPW{\Pchn}{W}}}$
are small enough for \Cref{eq:rate_redundancy} to hold. 
\end{IEEEproof}

\begin{lemma}
    \label{lemma:UEP_error2}
    For a \DMC, there exists a sequence of UEP codes with $k_n = \polyn$
    classes of messages,
    $\cA_i \in \TypeSetn{\Pchn^{(i)}}$, and
    rates $\hvect[k_n]{R}$,
    where $R_i \leq \Entropy{\Pchn^{(i)}} - \eta_n$,
    \begin{equation}
        \label{eq:def_etan}
        \eta_n \defeq \frac{2}{n}\left(
            \cardinality{\cX}^2 + \log(n+1) + \log k_n + 1
        \right),
    \end{equation}
    such that for any given $\bx \in \cA_i, i \in \Set{\hints[k_n]}$,
    %and its corresponding channel output $\by$,
    any  $\bx' \neq \bx$
    and $\bx' \in \cA_{i'}$, $i' \in \Set{\hints[k_n]}$, and
    any $\gamma \in \reals$,
    \begin{equation*} 
        \begin{split} &
            {\Prob{\MIPW{\Pchn^{(i')}}{P_{\bY|\bx'}} - R_{i'} \geq \gamma, \bY \sim \chinn{W}{\bx}}} \leq 
            \\ &
            (n+1)^{\cardinality{\cX}^2\cardinality{\cY}}
            \exp\left\{ -n\left[ \posfunc{R_{i'} + \gamma - \eta_n} - R_{i'} \right] \right\}. 
        \end{split}
        \label{eq:exp_decay}
    \end{equation*}
\end{lemma}
\begin{IEEEproof}[Proof sketch for \Cref{lemma:UEP_error2}]
    This proof is based on the coding scheme in Lemma 6 of~\cite{CsiszarLosslessJointExponent}.  In
    that construction, given channel conditional type $V$, the fraction of the output sequences
    correspond to $\cA_{i'}$ that overlaps with the output sequences of another codeword $\bx$ in a
    message set $\cA_i$ decays exponentially with the empirical mutual information
    $\MIPW{\Pchn^{(i)}}{V}$. Then by using a decoder based on empirical mutual information and by
    bounding the size of the output sequences that cause errors for the empirical mutual information
    decoder, we can show the desired result.
\end{IEEEproof}

The detailed proofs of \Cref{lemma:rate_redundancy,lemma:UEP_error2} are given in \Cref{sec:UEP_proofs}.
Below we present the proof for \Cref{thm:UEP_achievable}.

\begin{IEEEproof}[Proof of \Cref{thm:UEP_achievable}]
Fix some codeword length $n$. Without loss of generality, assume that the message is $m = (i,j)$ in
class $i$, which is mapped to a channel input $\bx(i,j) \in \cA_i$. Each codebook $\cA_i$ is drawn
uniformly over the type class of $\Pchn^{(i)} \in \PDSpaceXN$, where $\Pchn^{(i)}$ relates to
$\Pch^{(i)}$ (which is a general probability distribution that is not necessarily in $\PDSpaceXN$) by
\begin{equation*}
    |\Pchn^{(i)}(x) - \Pch^{(i)}(x)|_\infty \leq \frac{1}{n}.
\end{equation*}

For any $\by \in \cY^n$, define the measure for message $m$:
\begin{equation*}
    a_m(\by) \defeq I(\Pchn^{(i)}, P_{\by|\bx(i,j)}) - R_i,
    \label{eq:mi_decoder}
\end{equation*}
and let the decoder mapping $g_{C;n}: \cY^n \rightarrow \cM$ be defined as follows, using thresholds
$\gamma_n$ to be specified:
\begin{equation*}
    g_{C;n}(\by) = \begin{cases}
        m & %\text{ if }
        a_m(\by) \geq \gamma_n >  \max_{m'\neq m} a_{m'}(\by)
        \\
        \emptyset & \text{o.w. (declares a decoding failure)}
    \end{cases}
    %\label{}
\end{equation*}

The error event is the union of the following two events:
\begin{align}
\label{eq:E_1}
    \cE_1
    &= \left\{ 
    a_m(\by) < \gamma_n
    \right\}
    \\
    \label{eq:E_2}
    \cE_2
    &= \left\{ 
        \exists m'\neq m \in \cM \text{ s.t. }
        %a(\bx, \by) \geq \gamma_n,
        a_{m'}(\by) \geq \gamma_n
    \right\}
    .
\end{align}
Let $m'=(i',j')$ be a generic codeword different from $m$. For simplicity, we denote $\bx(i,j)$ and $\bx(i',j')$ by $\bx$ and
$\bx'$ respectively in the rest of the proof. Note that $i'$ may be equal to $i$.

We now choose
\begin{equation}
    \gamma_n = 2\eta_n + \frac{1}{2n}\log k_n + \frac{a}{n}\log n,
    \label{eq:gamma_n}
\end{equation}
where $\eta_n$ is defined by \Cref{eq:def_etan} in \Cref{lemma:UEP_error2} and
$a=\nicefrac{(d+1)}{2}$, where $d$ is the degree of the polynomial $k_n$. Note that
$\gamma_n=O\left(\nicefrac{\log n}{n}\right)$. \Cref{lemma:UEP_error2} shows
\begin{equation*} 
    \begin{split}
    \Prob{\cE_2}
    &=
    \sum_{j} \Prob{{\MIPW{\Pchn^{(i')}}{P_{\bY|\bx'}}}  - R_{i'} \geq \gamma_n, \bx' \in\cA_{i'},
    \bY \sim \chin{W}{\bx}}
    \\
    &\leq
 k_n
    (n+1)^{\cardinality{\cX}^2\cardinality{\cY}} \\
    & \quad   \exp\left\{ -n\min_{i'} \left[ \posfunc{R_{i'} + \gamma_n - \eta_n} - R_{i'} \right] \right\}
    \\
    &\leq k_n (n+1)^{\cardinality{\cX}^2\cardinality{\cY}} \\
    & \quad \exp\left\{ -n\left[ \eta_n + \frac{1}{2n}\log k_n + \frac{\log n^a}{n}  \right] \right\}
    \\
    &= \frac{\sqrt{k_n}}{n^a} = \BigO{\frac{1}{\sqrt{n}}}.
    \end{split} 
\end{equation*}
To analyze $\cE_1$, let
\begin{equation} 
    \label{eq:UEP_rates}
    \Delta R_i = \MIPW{\Pch^{(i)}}{W} - R_i - \gamma_n.
\end{equation}
Note that $  \Prob{\cE_1}$ may be written as
\begin{equation*}
    \Prob{ \MIPW{\Pchn^{(i)}}{P_{\bY|\bx}} -  \MIPW{\Pch^{(i)}}{W} \leq - \Delta R_i, \bY \sim \chinn{W}{\bx}  }.
    \label{eq:UEP_cE_1}
\end{equation*}
Now employing \Cref{eq:rate_redundancy_eps2} in
\Cref{lemma:rate_redundancy} with $\eps=\eps_i$ and 
\[ 
\delta_n = -    \Prob{\cE_2}  = \BigO{\frac{1}{\sqrt{n}}},
\]
we have 
\[
\Delta R_i = \sqrt{\frac{V(\Pch^{(i)},W)}{n}} \Qinv{\eps} + \BigO{\frac{\log n}{{n}}}
\]
is achievable.
By the union bound, the error probabilities are no more than
$\eps_i$, as required. Finally, \Cref{eq:UEP_rates} leads to 
\[
    R_i
    = \MIPW{\Pch^{(i)}}{W} - \sqrt{\frac{V_i}{n}}Q^{-1}(\eps_i) + \BigO{\frac{\log n}{n}}.
\]

\end{IEEEproof}

\section{Main Result: JSCC Dispersion}
\label{sec:JSCC}
We now utilize the UEP framework in \Cref{sec:UEP} to arrive at our main result. 

For the sake of investigating the finite block-length behavior,  we consider the \emph{excess
distortion} event $\ED$ defined in \Cref{eq:ED}.
When the distortion level is held fixed, Csisz\'ar gives lower and upper bounds on the exponential
decay of the excess distortion probability \cite{CsiszarJointExponent}. In this work, we fix the
excess distortion probability to be constant with the blocklength $n$
\begin{align}
    \label{eq:eps} 
    \Prob{\ED} = \eps
\end{align} 
and examine how the distortion thresholds $D_n$ approach the OPTA $D^*$ (the distortion achieving
equality in \Cref{eq:RDC}), or equivalently, how $R(P, D_n)$ approaches $R(P, D^*) = \rho C(W)$.
We find that it is governed by the joint source-channel dispersion
\Cref{eq:dispersion_formula}. In this formula, the source dispersion is given by
\cite{AmirYuvalDCC}:
\begin{align} 
    \label{eq:V_S} 
    V_S(P,D) & = 
    \Var{
        \left. \frac{\partial}{\partial Q_i} R(Q,D) \right|_{Q=P} 
    },
\end{align}
and the channel dispersion $V_C(W)$ is given by $\Vmin(W)$, which is assumed to be equal to $\Vmax(W)$.
%\footnote{See in the sequel for the case where they are not.}

 \begin{theorem} \label{thm:JSCC_discrete}
Consider a JSCC problem with a DMS $(\cS,\hat\cS,P,d)$, a DMC $(\cX,\cY,W)$ and bandwidth expansion
factor $\rho$. Let the corresponding OPTA be $D^*$. Assume that $R(Q,D)$ is differentiable w.r.t.
$D$ and twice differentiable w.r.t. $Q$ in some neighborhood of $(P,D^*)$. Also assume that the
channel dispersion $\Vmin(W)=\Vmax(W)>0$. Then for a fixed excess distortion probability $0<\eps<1$,
the optimal distortion thresholds $D_n$ satisfy:
\begin{equation}\label{eq:JSCC_bound}
    R(P,D_n) = \rho\cdot C(W) - \sqrt\frac{V_J(P,W,\rho)}{n} Q^{-1}(\eps) + \BigO{\tfrac{\log n}{n}}, \nonumber
\end{equation}
where $V_J(P,W,\rho)$ is the JSCC \rem{joint source-channel} dispersion \Cref{eq:dispersion_formula}.
\end{theorem}

\begin{figure}[t]
    \begin{center}
	        \begin{tikzpicture}[scale=0.8]
            %\draw [help lines] (0,0) grid (10,10);
            \begin{axis}[
                thick,
                width=4in,
                ymin=0, ymax=4.1,
                xmin=0, xmax=4.1,
                ylabel={$\rho I(\Phi,W)$},
                xlabel={$R(Q,D)$},
                ylabel style={at={(0.15,1.08)},rotate=-90},
                xlabel style={at={(1.10,0.15)}},
                ytick={3},
                xtick={0,1.5},
                xticklabels={$0$,$R(P\comma D)$},
                yticklabels={$C$},
                ]
                \addplot[no markers, gray!50, fill=gray!50, line width=0pt] coordinates
                    {
                    (0,0)
                    (3.8,3.8)
                    (3.8,0.02)
                    };
                    %\closedcycle;
                \addplot[thin, no markers, dashed] coordinates
                {
                    (1.5,3) (0,3)
                };
                \addplot[thin, no markers, dashed] coordinates
                {
                    (1.5,3) (1.5,0)
                };
                \addplot[mark=*] coordinates
                {
                    (1.5,3) 
                };
            \coordinate (Point) at (axis cs:1.5,3);
            \draw[thin] (Point) ellipse (2cm and 1.2cm);
            \draw[decorate,decoration={brace,raise=4pt},thick] 
                (axis cs:0.56,3) to node[midway,above=5pt] (bracket) {$\sqrt{{V_s}/{n}}$} (axis cs:1.5,3);
            \draw[decorate,decoration={brace,raise=4pt},thick] 
                (axis cs:1.5,3) to node[midway,right=5pt] (bracket) {\hspace{-2mm} $\sqrt{{\rho V_c}/{n}}$} (axis cs:1.5,2.31);
            
            \end{axis}
            %\draw (2,3) circle [radius=1.5cm];
        \end{tikzpicture}
	\end{center}
	\vspace{-2mm}
    \caption{Heuristic view of the main JSCC excess distortion event. The ellipse denotes the
    approximate one-standard-deviation region of the source-channel pair, while the gray area
    denotes the set of source-channel realizations leading to excess distortion.}
	\label{fig:jscc}
	\vspace{-2mm}
	\end{figure}
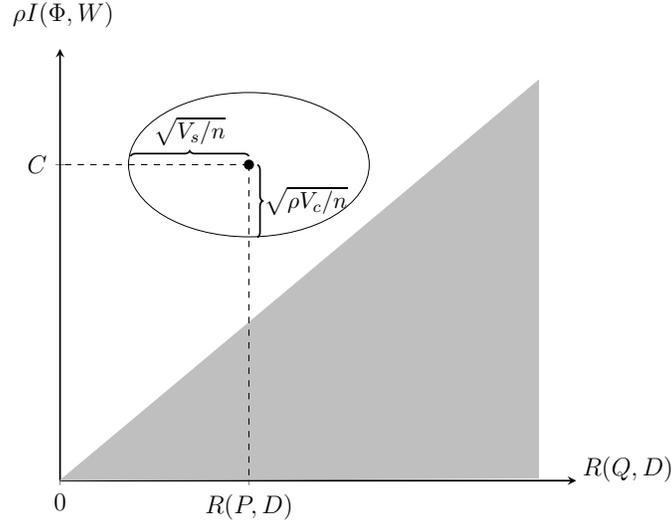

We can give a heuristic explanation to this result, graphically depicted in \Cref{fig:jscc}. We know
that the rate needed for describing the source is approximately Guassian, with mean $R(P,D_n)$ and
variance $V_S(P,D_n)/n$. Similarly, the mutual information supplied by the channel is approximately
Gaussian, with mean $\rho C(W)$ and variance $\rho V_C(W)/n$. We can now construct a codebook
per source type, and map this set of codebooks to a channel UEP code. According to \Cref{sec:UEP},
the dispersion of UEP given the rate of the chosen codebook is the same as only having that
codebook. Consequently, an error occurs if the source and channel empirical behavior
$(P_\bs,P_{\by|\bx})$ is such that \[ R(P_\bs,D_n) > \rho\cdot I(\Phi,P_{\by|\bx}). \] 
The difference between the left
and right hand sides is the difference of two independent approximately-Gaussian random variables,
thus is approximately Gaussian with mean $R(P,D_n)-\rho C$ and variance $V_J(P,W,D)$, yielding
\Cref{eq:JSCC_bound} up to the correction term.  However, for the proof we need to carefully
consider the deviations from Gaussianity of both source and channel behaviors. 

\begin{remark} 
    In the (rather pathological) case where $\Vmin(W)\neq\Vmax(W)$, we cannot draw anymore the
    ellipse of \Cref{fig:jscc}. This is since the variance of the channel mutual information will be
    different between codebooks that have error probability smaller or larger than $1/2$. We can use
    $\Vmin$ and $\Vmax$ for upper and lower bounds on the JSCC dispersion.  Also, when $\eps$ is
    close to zero or one, the dispersion of the channel part is very well approximated by
    $\Vmin$ or $\Vmax$, respectively. 
\end{remark}

\begin{remark} 
    The source and channel dispersions are known to be the second derivatives (with respect to the
    rate) of the source exponent at rate $R(P,D)$ and of the channel exponent at rate $C(W)$,
    respectively. Interestingly, the JSCC dispersion \Cref{eq:dispersion_formula} is also connected
    to the second derivative of the JSCC exponent \cite{CsiszarJointExponent}:
    \[ E(P,W,D,\rho) {\,=} \min_{R(P,D)\leq R \leq C} [E_S(P,D) + \rho E_C(W)] \] 
    (where $E_S$ and $E_C$ are the lossy source coding and sphere-packing exponents%
    \footnote{Sphere-packing exponent is only achievable when $R$ is close to $C$, but this is
    sufficient for the derivative at $R=C$.}%
    , respectively) via
    \begin{equation*} 
    V_J(P,W,\rho) = 
    \left[ \frac{\partial^2 E(P,W,D,\rho)}{\partial R(P,D)^2} \bigg|_{D=D^*(P,W,\rho)}\right]^{-1},
    \end{equation*}
    where in the derivative $P$ is held fixed.
\end{remark}  

The achievability part of \Cref{thm:JSCC_discrete} relies on the following lemma. 
\begin{lemma}[JSCC Distortion Redundancy]\label{lemma:JSCCRedundancy} 
    Consider a JSCC problem with a DMS $(\cS,\hat\cS,P,d)$, a DMC $(\cX,\cY,W)$ and bandwidth
    expansion factor $\rho$. Let $n$ be the length of the source block length, and let
    $m\defeq \floor{\rho n}$ be the length of the channel block length. 
    Let $\Phi$ be an arbitrary
    distribution on $\cX$, and let $\Phi_m \in \cP_m(\cX)$ be an $m$-type that approximates $\Phi$ as  
    \[ \|\Phi - \Phi_m\|_\infty\leq \frac{1}{m}. \] 
    Let the channel input $\bx \in \cX^m$ have type $\Phi_m$. 
    Further, let $D^*(\Phi)$ be the solution to $R(P,D(\Phi))=\rho I(\Phi,W)$. 
    Assume that $R(Q,D)$, the RDF of a
    source $Q$ with the same distortion measure, is twice differentiable w.r.t. $D$ and the elements
    of $Q$ at some neighborhood of $(P,D^*(\Phi))$.  Let $\eps$ be a given probability and let
    $D_n>0$ be the infimal value s.t.
    \begin{equation}\label{eq:def_DeltaDn}
        %\Prob{R(P_\bS,D_n) > \rho I(\Phi_m,P_{\bY|\bx}) } \leq \eps.
        \Prob{R(P_\bS,D_n) > \rho I(\Phi_m,P_{\bY|\bx}) } = \eps.
    \end{equation}
    Then, as $n$ grows,
    \begin{equation}
        \label{eq:DistortionRedundancyJSCC}
        \begin{split}
        R(P,D_n) = \rho I(\Phi,W) 
        & - \sqrt\frac{V_S(P) + \rho V_C(\Phi,W)}{n}Q^{-1}(\eps) 
        \\ & + O \left(\frac{\log n}{n}\right). 
        \end{split}
    \end{equation}
    In addition, for any channel input (i.e., $\Phi_m$ is not restricted and may also depend upon the source sequence),
    \begin{equation}
        \label{eq:DistortionRedundancyJSCC_unrestricted}
        R(P,D_n) \leq \rho C(W) - \sqrt\frac{V_J(P,W,\rho) }{n}Q^{-1}(\eps)  + O \left(\frac{\log n}{n}\right), 
    \end{equation} where $V_J(P,W,\rho)$ is given by \Cref{eq:dispersion_formula}.
    Furthermore, all the above holds even if replace \Cref{eq:def_DeltaDn} with
    \begin{equation}
        \label{eq:def_DeltaDn_pert}
        \Prob{R(P_\bS,D_n) > \rho I(\Phi_m,P_{\bY|\bx}) + \xi_n} 
        =
        \eps + \zeta_n,
    \end{equation}
    for any given (vanishing) sequences $\xi_n,\zeta_n$, as long as 
    $\xi_n = O\left(\frac{\log n}{n}\right)$ 
    and 
    $\zeta_n = O\left(\frac{\log n}{\sqrt n}\right)$.
\end{lemma}
\begin{IEEEproof}[Proof sketch for \Cref{lemma:JSCCRedundancy}]
    Similar to \Cref{lemma:rate_redundancy}, we apply Taylor expansion to $R(P_\bS,D_n)$ and show
    that the first order term again can be expressed as sum of $n$ independent random variables, and
    neglecting higher order terms does not affect the statement.  
    Then $R(P_\bS,D_n) - \rho I(\Phi_m,P_{\bY|\bx})$ can be shown to be the sum of $n+m$ indenpdent
    random variables, with total variance essentially $(V_S + \rho V_C(\Pch, W))/n$. Finally,
    similar to the derivation in \Cref{lemma:rate_redundancy}, we apply the Berry-Esseen theorem
    and show \Cref{eq:DistortionRedundancyJSCC} and \Cref{eq:def_DeltaDn_pert} are true.
\end{IEEEproof}

The detailed proof of \Cref{lemma:JSCCRedundancy} is given in \Cref{sec:JSCC_proofs}.

The converse part of \Cref{thm:JSCC_discrete} builds upon the following result, which states
that for any JSCC scheme, the excess-distortion probability must be very high if the empirical
mutual information over the channel is higher than the empirical source RDF.

\begin{lemma}[Joint source channel coding converse with fixed types]
    \label{lemma:jscc_converse_given_types}
    For a JSCC problem, given a source type $Q \in \PDSpaceN{\cS}$ 
    and a channel input type $\Pch \in \PDSpaceXN$, 
    let $G(Q, \Pch)$ be the set of source seqeuences in
    $\TypeSetn{Q}$ that are mapped (via JSCC encoder $f_{J;n}$) to channel codewords with type
    $\Pch$, \ie, 
    \[
    G(Q, \Pch) \defeq \Set{\bs \in \Types_Q^n : \bx = f_{J;n}(\bs) \in T_\Phi^n}.
    \]
    Define all the channel outputs that covers $\bs$ with distortion $D$ as $\hB(\bs, D)$, \ie, 
    \begin{equation}
        \hB(\bs, D) = \SetDef{\by \in \cY^m}{ d(\bs, g_{J;n}(\by)) \leq D}
        \label{eq:s_cover}
    \end{equation}
    where $m = \floor{\rho n}$ and $g_{J;n}$ is the JSCC decoder. 
    If 
    \begin{equation}
        \cardinality{G(Q, \Pch)} 
        \geq 
        \frac{1}{(n+1)^{\cardinality{\cX}+1}} \cardinality{\TypeSetn{Q}}, 
    \end{equation}
    then for a given distortion $D$ and
    a channel with constant composition conditional distribution $V \in \cP_m(\cY|\Pch)$, we have
    \begin{equation}
        \frac{1}{\cardinality{G(Q, \Pch)}}
        \sum_{\bs_i \in G(Q, \Pch)}
        \frac{ 
            \cardinality{ 
                \TypeShelln[m]{V}{f(\bs_i)} 
                \cap 
                \hB(\bs_i, D)
                }
        }
        {
        \cardinality{\TypeShelln[m]{V}{f(\bs_i)}}
        }
        \leq
        p(n)
        \exp^{
            - n
            \left[ 
            R(Q,D) 
            -
            \rho I(\Pch, V)
            \right]^+
        }
        \label{eq:converse_given_types}
    \end{equation}
    where $p(n)$
    is a polynomial that depends only on the source, channel and reconstruction alphabet sizes and
    $\rho$.
\end{lemma}
The detailed proof of \Cref{lemma:jscc_converse_given_types} is given in \Cref{sec:JSCC_proofs}.
The proof uses an approach similar to that in the strong channel coding converse~\cite{DueckKorner}.
%and bound the no-excess distortion probability via \Cref{lemma:jscc_converse_given_types}.

Below we present the proof for \Cref{thm:JSCC_discrete}. The achievability proof is based on
\Cref{lemma:JSCCRedundancy,lemma:UEP_error2}, where we do not use directly
\Cref{lemma:rate_redundancy} or \Cref{thm:UEP_achievable}, thus we do not suffer from the
non-uniformity problem (see \Cref{rem:non_uniform}). In other words, rather than evaluating the
error probability per UEP codebook, we directly evaluate the average over all codebooks.

\begin{IEEEproof}
\textbf{Achievability:} 
Let \begin{equation} \label{eq:k_n} k_n = (n+1)^{|\cS|+1} = \polyn. \end{equation}
At each block length $n$, we construct a source code $\cC=\{\cC_i\}$ as follows (the index $n$ is
omitted for notational simplicity). Each code $\cC_i$ corresponds to one type $Q_i \in (\PDSpaceXN \bigcap \Omega_n)$, where
\begin{align*}\label{eq:CoveredTypes}
   \Omega_n = \left\{Q: \|P-Q\|_2^2 \leq |\cS| \frac{\log n}{n} \right\}. 
\end{align*}
According to the refined type-covering Lemma \cite{YuSpeedCovering}, there exists codes $\cC_i$ of rates
\begin{equation}
    \label{eq:source_rates}
    R_i \leq R(Q_i,D_n) + O\left(\frac {\log n}{n}\right),
\end{equation}
that completely $D_n$-cover the corresponding types (where the redundancy term is uniform). We
choose these to be the rates of the source code. The chosen codebook and codeword indices are then
communicated using a dispersion-optimal UEP scheme as described in \cref{sec:UEP} with a
capacity-achieving channel input distribution $\Phi\in\Pi(W)$. Specifically, each source codebook is
mapped into a channel codebook of block length $\floor{\rho n}$ and rate 
\begin{equation}
    \tilde R_i = \frac{R_i}{\rho},
    \label{eq:channel_rates}
\end{equation}
as long as 
\begin{equation} 
    \label{eq:mapping_threshold} 
    \tilde R_i \leq H(\Phi) - \eta_n, 
\end{equation} 
where $\eta_n$ is defined in \Cref{lemma:rate_redundancy}. Otherwise, the mapping is arbitrary and
we assume that an error will occur. The UEP scheme is thus used with different message classes at
each $n$; such a scheme can only perform better than a scheme where the message classes accumulate,
see \Cref{rem:cumulative}, thus we can use the results of \Cref{sec:UEP} with number of codebooks:
\[   
\sum_{n'=1}^n |\PDSpaceXN\bigcup\Omega_{n'}| \leq n \cdot |\PDSpaceXN| \leq (n+1)^{|\cS|+1} = k_n,
\]
where $\PDSpaceXN$ is defined in \Cref{eq:all_types}.

Error analysis: an excess-distortion event can occur only if one of the following events happened: 
\begin{enumerate} 
    \item $P_\bs \notin \Omega_n$, where $P_\bs$ is the type of $\bs$. 
    \item $\tilde R_i \geq H(\Phi) - \eta_n$.  
    \item $\cE_2$ \Cref{eq:E_2}: an unrelated channel codeword had high empirical mutual information. 
    \item $\cE_1$ \Cref{eq:E_1}: the true channel codeword had low empirical mutual information. 
\end{enumerate}
We show that the first three events only contribute to the correction term.
According to \cite[Lemma 2]{AmirYuvalDCC},  \[ \Prob{P_\bs \notin \Omega_n} \leq \frac{2|\cS|}{n^2}. \]
By our assumption on the differentiability of $R(P,D)$, for large enough $n$ the second event does not happen for any type in $\Omega_n$. 
By \Cref{lemma:UEP_error2}, the probability of the third event is at most $O(1/\sqrt{n})$, uniformly. 
Thus, by the union bound, we need the probability of the last event to be at most $\eps_n = \eps - O(1/\sqrt{n})$. 

Now following the analysis of $\cE_1$ in the proof of \Cref{thm:UEP_achievable}, 
\Cref{eq:UEP_rates} indicates that event $\cE_1$ is equivalent to
\[ \MIPW{\Pch}{P_{\bY|\bx}} \leq \tilde R_i + \gamma_n \]
where $\gamma_n$ is defined in \Cref{eq:gamma_n}.
\Cref{eq:source_rates} and \Cref{eq:channel_rates} indicates this is equivalent to 
\[ 
\rho \MIPW{\Pch}{P_{\bY|\bx}} \leq R(P_\bs,D_n) + O\left(\frac{\log n}{n}\right). 
\] 
On account of \Cref{lemma:JSCCRedundancy}, this can indeed be satisfied with $\eps_n$ as required.  

\textbf{Converse:} 
At the first stage of the proof we suppress the dependence on the block length $n$ for conciseness.
We first lower-bound the excess-distortion probability given that the source type is some $Q \in \PDSpaceN{S}$.

Let $\alpha(Q,\Phi) \triangleq \Prob{P_\bX = \Phi | P_\bS = Q}$ be
the probability of having input type $\Phi$ giving that the source type is $Q$.
Noting that given a source type, all strings within a type class are equally likely, we have
\begin{align*}
\alpha(Q,\Phi) 
= \frac{|\{\bs \in \Types_Q^n : \bx = f_{J;n}(\bs) \in T_\Phi^n\}|} {|\Types_Q^n|} 
= \frac{\abs{G(Q, \Pch)}}{|\Types_Q^n|}.
\end{align*}
Now we have
\[ 
\Prob{\ED|P_\bS=Q} = \sum_{\Phi \in \PDSpaceXN} \alpha(Q,\Phi) \Prob{\ED|P_\bS=Q,P_\bX=\Phi}.
\]
Define the class of ``frequent types'' based on $\alpha(Q, \Pch)$:
\[
A(Q) \triangleq \left\{\Phi\in\PDSpaceXN : \alpha(Q,\Phi) \geq \frac{1}{(n+1)^{|\cX|+1}} \right\}. 
\]
Note that 
\[
\Prob{ P_\bX  \notin A(Q)| P_\bS = Q} 
\leq |\PDSpaceXN| \frac{1}{(n+1)^{|\cX|+1}} 
\leq \frac{1}{n+1}, 
\] thus $A(Q)$ is nonempty.
Trivially, we have:
\begin{align*} 
    \Prob{\ED | P_\bS=Q} 
    & \geq \sum_{\Phi \in A(Q)} \alpha(Q,\Phi) \Prob{\ED|P_\bS=Q,P_\bx=\Phi } 
    \\ 
    &= \sum_{\Phi \in A(Q)} \alpha(Q,\Phi) \sum_{V \in\PDSpaceN{\cY|\Phi}} \Prob{P_{\by|\bx}=V|P_\bx=\Phi}\Prob{\ED|P_\bS=Q,P_\bx=\Phi,P_{\by|\bx}=V}. 
\end{align*}
Next we use \Cref{lemma:jscc_converse_given_types} to assert, for all $\Phi \in A(Q)$:
\begin{align*}  
    \Prob{\ED|P_\bS=Q,P_\bx=\Phi,P_{\by|\bx}=V }
    & \geq 1 - 
        \frac{1}{\cardinality{G(Q, \Pch)}}
        \sum_{\bs_i \in G(Q, \Pch)}
        \frac{ 
            \cardinality{ 
                \TypeShelln[m]{V}{f(\bs_i)} 
                \cap 
                \hB(\bs_i, D)
                }
        }
        {
        \cardinality{\TypeShelln[m]{V}{f(\bs_i)}}
        }
    \\ 
    & \geq  1 - p(n) \exp\{-n[R(Q,D)-\rho I(\Phi,V)]\},
\end{align*} 
where $p(n)$ is given in \Cref{eq:poly_term}.
Since $\sum_{\Phi\in A(Q)} \alpha(Q,\Phi) \leq 1$, we further have:
\begin{align*}
    \Prob{\ED|P_\bS=Q} 
    \geq\;
    &
    1 - \frac{1}{n+1}
    \\
    & +
    p(n) \sum_{V\in \PDSpaceN{\cY|\Phi^*(Q)}}  \Prob{P_{\by|\bx}=V|P_\bx=\Phi^*(Q)} \exp\{-n[R(Q,D)-\rho I(\Phi^*(Q),V)]\}, 
\end{align*} 
where $\Phi^*(Q)$ minimizes  the expression over all $\Phi\in A(Q)$ (if there are multiple
maximizers, it is chosen arbitrarily). Collecting all source types, we have: 
\begin{align*} 
    \Prob{\ED} 
    \geq 
    \frac{n}{n+1} -  p(n) \sum_{Q\in\PDSpaceN{\cS}} \sum_{V\in \PDSpaceN{\cY|\Phi^*(Q)}} 
    & \Prob{P_\bS=Q} \Prob{P_{\bY|\bX}=V|P_\bX=\Phi^*(Q) } \cdot
    \\ 
    & \exp\{-n[R(Q,D)- \rho I(\Phi^*(Q),V)]\}. 
\end{align*} 
At this point we return the block length index $n$. Let $\Delta_n$ be some vanishing sequence to be specified later.
Define the set 
\[ 
B(\Delta_n) \triangleq \{Q\in\PDSpaceN{\cS}, V \in \PDSpaceN{\cY|\Phi^*_n(Q)} : R(Q,D) - I(\Phi^*_n(Q),V) > \Delta_n\}. 
\]
For any sequence $\Delta_n$ we can write:  
\[ \Prob{\ED} \geq  \Prob{B(\Delta_n)} \left[ \frac{n}{n+1} - p(n) \exp\{-n\Delta_n\} \right], \]
where $\Prob{B(\Delta_n)} = \Prob{\bS: P_{f_{J;n}(\bS)} \in B(\Delta_n)}$. 
Now choose $n \Delta_n = (1 + p(n)) \log(n+1)$ to obtain:
\[   
\Prob{\ED} 
\geq 
\left(1-\frac{2}{n+1}\right) \Prob{B(\Delta_n)} 
\geq 
\frac { \Prob{B(\Delta_n)}} {1+\frac{2}{n-1}}. 
\]
Since we demand that $ \Prob{\ED} \leq \eps$ for all $n$, and inserting the definition of
$B(\Delta_n)$ it must be that 
\[ \Prob{R(P_\bS,D) - \rho I(\Phi^*_n(P_\bS),V) > \Delta_n} \leq \eps \left(1+\frac{2}{n-1}\right). \] 
Seeing that $\Delta_n = O\left(\nicefrac{\log n}{n}\right)$, the desired result follows on account
of \Cref{eq:DistortionRedundancyJSCC_unrestricted} in \Cref{lemma:JSCCRedundancy}.
\end{IEEEproof}

\section{The Loss of Separation}
\label{sec:separation}
In this section we quantify the dispersion loss of a separation-based scheme with respect to the
JSCC one. Using the separation approach, the interface between the source and channel parts is a
fixed-rate message, as opposed to the variable-rate interface used in conjunction with multiple
quantizers and UEP, shown in this work to achieve the JSCC dispersion.

Formally, we define a separation-based encoder as the concatenation of the following elements. 
\begin{enumerate} 
    \item A source encoder $f_{S;n}: \cS^n \rightarrow \cM_n$. \item A source-channel mapping
        $\cM_n \rightarrow \cM_n$. 
    \item A channel encoder $f_{C;n}: \cM_n \rightarrow \cS^{\floor{\rho n}}$. 
\end{enumerate} 

The interface rate is $R_n = \log{\cardinality{\cM_n}} / n$. Finally, the source-channel mapping is
randomized, in order to avoid ``lucky'' source-channel matching that leads to an effective ``joint''
scheme.\footnote{For instance, the UEP scheme could be presented as a separation one if not for the
randomized mapping.} We assume that it is uniform over all permutations of $\cM_n$, and that it is
known at the decoder as well. Consequently, the decoder is the obvious concatenation of elements in
reversed order. The excess distortion probability of the scheme is defined as the mean over all
permutations. 

In a separation-based scheme, an excess-distortion event occurs if one of the following: either the
source coding results in excess distortion, or the channel  coding results in a decoding error.
Though it is possible that no excess distortion will occur when a channel error occurs (whether the
source code has excess distortion or not), the probability of this event is exponentially small.
Thus at every block-length $n$, the excess-distortion probability $\eps$ satisfies
\begin{align} 
    \label{eq:probabilities_conv} 
    \eps = \eps_{S;n} * \eps_{C;n} - \delta_n 
\end{align} 
where $a*b=a+b-ab$, $\eps_{S;n}$ and $\eps_{C;n}$ are the source excess-distortion probability and
channel error probability, repectively, at blocklength $n$, and $\delta_n$ is exponentially
decaying with $n$. In this expression we take a fixed  $\eps$, in accordance with the dispersion setting; the
system designer is still free to choose $\eps_{S;n}$ and $\eps_{C;n}$ by adjusting the rates
$R_n$, as long as \Cref{eq:probabilities_conv} is maintained.

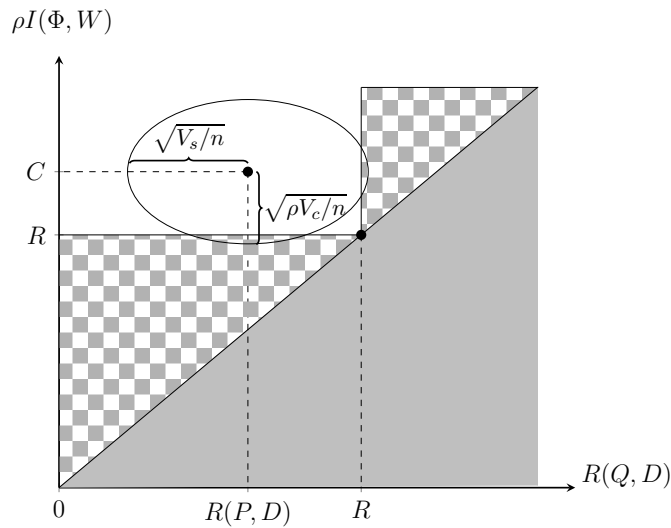
\begin{figure}[t]
    \begin{center}
                \begin{tikzpicture}[scale=0.8]
            %\draw [help lines] (0,0) grid (10,10);
            \begin{axis}[
                thick,
                width=4in,
                ymin=0, ymax=4.1,
                xmin=0, xmax=4.1,
                ylabel={$\rho I(\Phi,W)$},
                xlabel={$R(Q,D)$},
                ylabel style={at={(0.15,1.08)},rotate=-90},
                xlabel style={at={(1.10,0.15)}},
                ytick={2.4,3},
                xtick={0,1.5,2.4},
                xticklabels={$0$,$R(P\comma D)$,$R$},
                yticklabels={$R$,$C$},
                ]
                \addplot[no markers, gray!50, fill=gray!50, line width=0pt] coordinates
                    {
                    (0,0)
                    (3.8,3.8)
                    (3.8,0.02)
                    };
                    %\closedcycle;
                \addplot[thin, no markers, dashed] coordinates
                {
                    (1.5,3) (0,3)
                };
                \addplot[thin, no markers, dashed] coordinates
                {
                    (1.5,3) (1.5,0)
                };
                \addplot[mark=*] coordinates
                {
                    (1.5,3) 
                };
                \addplot[mark=*] coordinates
                {
                    (2.4,2.4) 
                };
                      \addplot[no markers, pattern=checkerboard,pattern color=black!30, line width=0pt] coordinates
                    {
                    (2.4, 2.4)
                    (3.8,3.8)
                    (2.4,3.8)
                    };
                  \addplot[no markers, pattern=checkerboard,pattern color=black!30, line width=0pt] coordinates
                    {
                    (2.4, 2.4)
                    (0,0)
                    (0,2.4)
                    };
          \addplot[thin, no markers] coordinates
                {
                    (2.4,3.8) (2.4,2.4) (0,2.4)
                };	
                     \addplot[thin, no markers, dashed] coordinates
                {
                    (2.4,2.4) (2.4,0)
                };
            \coordinate (Point) at (axis cs:1.5,3);
            \draw[thin] (Point) ellipse (2cm and 1.2cm);
            \draw[decorate,decoration={brace,raise=4pt},thick] 
                (axis cs:0.56,3) to node[midway,above=5pt] (bracket) {$\sqrt{{V_s}/{n}}$} (axis cs:1.5,3);
            \draw[decorate,decoration={brace,raise=4pt},thick] 
                (axis cs:1.5,3) to node[midway,right=5pt] (bracket) {\hspace{-2mm} $\sqrt{{\rho V_c}/{n}}$} (axis cs:1.5,2.31);
            
            \end{axis}
            %\draw (2,3) circle [radius=1.5cm];
        \end{tikzpicture}
    \end{center}
    \caption{Main JSCC excess distortion event: the loss of separation.}
    \label{fig:jscc_sep}
\end{figure}

\rem{\begin{align} R(P,D_n)  = R_n - \sqrt\frac{V_s(P,D)}{n} \Qinv{\eps_{s;n}} +O\left(\frac{\log n}{n}\right)\\ R_n  = \rho C(W) - \sqrt\frac{\rho V_c(W)}{n} \Qinv{\eps_{c;n}} + O\left(\frac{\log n}{n}\right) \end{align} It follows by \Cref{eq:probabilities_conv} }

We now employ the source and channel dispersion results \Cref{eq:S_dispersion},
\Cref{eq:C_dispersion}, which hold up to a correction term
$\BigO{\nicefrac{log(n)}{n}}$,\footnote{The redundancy terms are in general functions of the error
probabilities, but for probabilities bounded away from zero and one they can be uniformly bounded;
it will become evident that for positive and finite source and channel dispersions, this is indeed
the case.} to see that that for the optimal separation-based scheme: 
    \begin{equation} 
        \begin{split}  
            R(D_n) &= 
            \rho C(W)  
            - \min_{\eps_{S;n} * \eps_{C;n} \leq \eps}  
            \left[  \sqrt{\frac{V_s(P,D^*)}{n}} \Qinv{\eps_{S;n}} \right. 
            \\
            & \quad + \left. \sqrt{\frac{\rho V_c(W)}{n}} \Qinv{\eps_{C;n}} \right] 
            +O\left(\frac{\log n}{n}\right) . 
        \end{split}
    \end{equation} 
It follows, that up to the correction term it is optimal to choose \emph{fixed} probabilities
$\eps_{S;n}=\eps_S$ and $\eps_{C;n}=\eps_C$. Furthermore, the dependancy on $n$ is the same as in
the joint source-channel dispersion \Cref{eq:DistortionRedundancyJSCC}, but with different
coefficient for the $1/\sqrt{n}$ term, \ie, 
\begin{equation}
    R(D_n) = 
            \rho C(W)  
            - \sqrt{\frac{V_{\text{sep}}}{n}} \Qinv{\eps}
            +O\left(\frac{\log n}{n}\right). 
    \label{eq:RD_sep}
\end{equation}
Note that in the limit $\eps \rightarrow 0$, $\sqrt{V_{\text{sep}}} = \sqrt{V_S} + \sqrt{\rho V_C}$.

In order to see why separation must have a loss, consider \Cref{fig:jscc}. The separation scheme designer is free to choose the digital interface rate $R$. Now whenever the random source-channel pair is either to the right of the point $(R,R)$ due to a source type with $R(P_\bS,D)>R$, or below it due to channel behavior $I(\Phi,P_{\bY|\bx})<R$, an excess distortion event will occur. Comparing to optimal JSCC, this adds the chessboard-pattern area on the plot. The designer may optimize $R$ such that the probability of this area is minimized, but for any choice of $R$ it will still have a strictly positive probability. 

\begin{figure}[!t] 
\centering 
      \includegraphics[scale = 0.6]{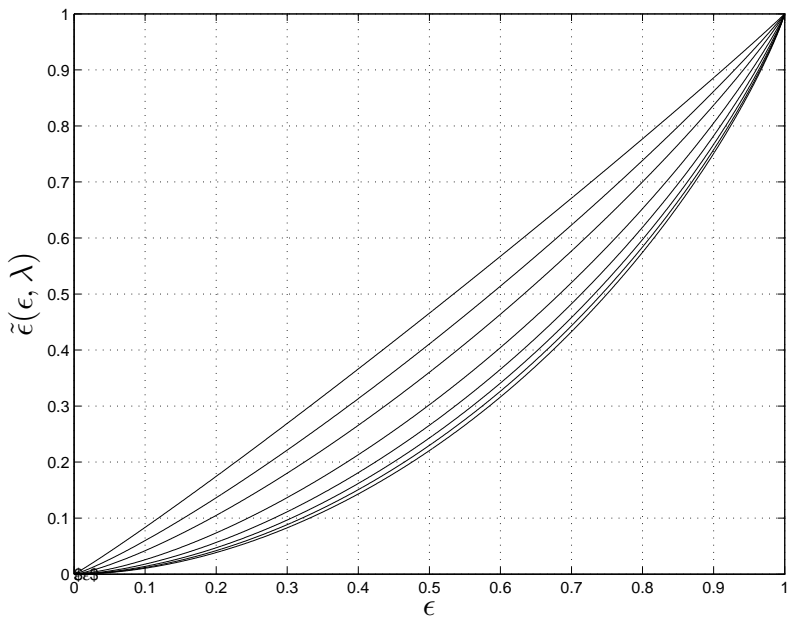}
 \caption{$\tilde\eps(\eps,\lambda)$ as a function of $\eps$ for different values of $\lambda$. From bottom to top curve, $\lambda=\{1,2,3,5,10,30,100,1000\}$.}
\label{fig:separation} 
\end{figure}

For quantifying the loss, it is tempting to look at the ratio between the coefficients of the
$1/\sqrt{n}$ terms. However, this ratio may be in general infinite or negative, making the
comparison difficult. We choose to define the equivalent probability $\tilde\eps$ by rewriting 
\Cref{eq:RD_sep} as
\begin{align}
    \label{eq:tilde_eps_def} 
    R(D_n) = \rho C  - \sqrt\frac{V(P,D,W,\rho)}{n} \Qinv{\tilde\eps} +
    O\left(\frac{\log n}{n}\right). 
\end{align} 
Thus, $\tilde\eps<\eps$ is the excess-distortion probability that a JSCC scheme could achieve under
the same conditions, when the seperation scheme achieves $\eps$. Substitution reveals that
\begin{align} 
    \label{eq:tilde_eps} 
    \tilde\eps(\eps,\lambda) = 
    Q\left( \frac{ \min_{\eps_{s} * \eps_{c} \leq \eps} \Bigl[ \Qinv{\eps_s} + \sqrt{\lambda} \Qinv{\eps_c}  \Bigr]}{\sqrt{1+\lambda}}\right), 
\end{align} 
where 
\begin{align} \label{eq:lambda} \lambda \triangleq \frac{\rho V_c}{V_s}. \end{align} 
In general, numerical optimization is needed in order to obtain the equivalent probability. However,
clearly $\tilde\eps(\eps,\lambda) = \tilde\eps(\eps,1/\lambda)$. In the special symmetric case
$\lambda=1$ one may verify that the optimal probabilities are \[ \eps_s = \eps_c = 1 -
\sqrt{1-\eps}, \] thus 
\[
\tilde\eps(\eps,1) = Q \left(\sqrt{2} \Qinv{1-\sqrt{1-\eps}} \right). 
\] 
This reflects a large loss for low error probabilities. On the other hand,
\[\lim_{\lambda\rightarrow 0} \tilde\eps(\eps,\lambda) = \lim_{\lambda\rightarrow \infty}
\tilde\eps(\eps,\lambda) = \eps . \] It seems that the symmetric case is the worst for separation,
while when $\lambda$ grows away from $1$, either the source or the channel behave deterministically
in the scale of interest, making the JSCC problem practically a digital one, i.e., either source
coding over a clean channel or channel coding of equi-probable messages. This is somewhat similar to
the loss of separation in terms of excess distortion exponent. This behavior is depicted in
\Cref{fig:separation}.

\section{BW Expansion and Lossless JSCC}
\label{sec:lossless}
We now wish to change the rules, by allowing the BW expansion ratio $\rho$, which was hitherto
considered constant, to vary with the blocklength $n$. 
More specically, we takes some sequence $\rho_n$ with $\lim_{n\rightarrow\infty}\rho_n = \rho$.
It is not hard to verify that the results of
\Cref{sec:JSCC} remain valid%
\footnote{
Note that now the application of 
Berry-Esseen theorem is more involved, as 
we are now summing $\rho_n n + n$ independent random variables. However, its application still holds
and results in \Cref{sec:JSCC} can be proved by keeping track of $\rho_n$ explicitly.},
and
$\rho_n$ and $D_n$ are related via:
\begin{equation}\label{eq:DistortionJSCC_rho}
    R(P,D_n) = \rho_n C(W) - \sqrt\frac{V_J(P,W,\rho)}{n}Q^{-1}(\eps)  + O \left(\frac{\log n}{n}\right),
\end{equation} where for the calculation of the JSCC dispersion we use $D^*(P,W,\rho)$. In particular, one may choose to work with a \emph{fixed} distortion threshold $D=D^*(P,W,\rho)$, and then \Cref{eq:DistortionJSCC_rho} describes the convergence of the BW expansion ratio sequence to its limit $\rho$. 

Equipped with this, we can now formulate a meaningful lossless JSCC dispersion problem. In (nearly) lossless coding we demand $\hat{\bS} = \bS$, otherwise we say that an \emph{error event} $\mathcal{E}$ has occurred. We can see this as a special case of the lossy JSCC problem with Hamming distortion:
\[ d(s_i,\hat s_i) = \begin{cases} 1 & \hat s_i = s_i \\ 0 & \text{otherwise}, \end{cases} \] and with distortion threshold $D=0$. While this setting does not allow for varying distortion thresholds, one may be interested in the number of channel uses needed to ensure a fixed error probability $\eps$, as a function of the blocklength $n$. As an immediate corollary of \Cref{eq:DistortionJSCC_rho}, this is given by:
\begin{equation} 
    \label{eq:lossless_dispersion}  
    \rho_n = \frac{H(P)}{C(W)} +
    \sqrt\frac{V_J(P,W,\rho)}{n}\frac{Q^{-1}(\eps)}{C(W)} + O \left(\frac{\log n}{n}\right).
\end{equation} In lossless JSCC dispersion, the source part of $V_J(P,W,\rho)$ simplifies to $\Var{\log P}$, in agreement with the lossless source coding dispersion of Strassen \cite{Strassen62_Asymptotische}.

%\section{The Quadratic-Gaussian Case}
%\label{sec:Gaussian}

\section*{acknowledgement}

The authors would like to thank Fadi Alajaji, Uri Erez, Meir Feder, Ligong Wang and Gregory W.
Wornell for helpful discussions. 

\appendices
\def\thesubappendix{\arabic{subappendix}}
\renewcommand{\theequation}{\thesection.\arabic{equation}}

\label{sec:proofs}
% extra defs
%where it is not hard to show that
%\begin{equation}
%    \InfoVar{\Pch}{W} \geq \CondInfoVar{\Pch}{W}.
%    %\label{}
%\end{equation}
%
%   unconditional information variance
%       \begin{equation}
%           \InfoVar{\Pch}{W}
%           = \sum_{x \in \cX} \sum_{y \in \cY} \Pch(x) \chio{W}{y}{x}
%           \log^2\frac{\chio{W}{y}{x}}{\Pch W(y)} - [\MIPW{\Pch}{W}]^2
%           %\label{}
%       \end{equation}

%\section{Proofs}
%\subsection{Unconditional and conditional information variance}
%\label{sec:VltU}
%In this section we show the simple fact that for the same input distribution $\Pch$ and channel $W$,
%conditional information variance is smaller than the unconditional information variance.
%
%\begin{lemma}
%    Given $\Pch$ and $W$, $\CondInfoVar{\Pch}{W} \leq \InfoVar{\Pch}{W}$.
%    \label{lemma:VltU}
%\end{lemma}
%
%\begin{proof}
%Note that
%\begin{align*}
%    \CondInfoVar{\Pch}{W} + \Ep{\Pch}{\left[ \KLD{W_x}{\Pch W} \right]^2}
%    &= 
%    \InfoVar{\Pch}{W} + \left[ \MIPW{\Pch}{W} \right]^2
%\end{align*}
%and
%\begin{align*}
%    \left[  \MIPW{\Pch}{W} \right]^2 
%    &= \left[ \Ep{\Pch}{\KLD{W_x}{\Pch W}} \right]^2
%    \\
%    &\leq \Ep{\Pch}{ \left[\KLD{W_x}{\Pch W}\right]^2} 
%\end{align*}
%Hence $\CondInfoVar{\Pch}{W} \leq \InfoVar{\Pch}{W}$.
%\end{proof}

\section{Proofs for UEP channel coding dispersion}
\label[app]{sec:proofs_UEP}
In this appendix we provides proofs for results in \Cref{sec:UEP}. We start by analyzing the Taylor
expansion of empirical mutual information in \Cref{sec:proofs_empirical_MI}, which is crucial for
proving \Cref{lemma:rate_redundancy}, then we proceed to prove
\Cref{lemma:rate_redundancy,lemma:UEP_error2} in \Cref{sec:UEP_proofs}.

\subsection{Analysis of the empirical mutual information}
\label[app]{sec:proofs_empirical_MI}

In this section, we investigate the Taylor expansion of the empirical mutual information at expected
mutual information, \ie, 
\begin{align}
    \label{eq:empirical_mi_taylor}
    I(\Phi,P_{\bY|\bx}) = I(\Phi,W) &+ \sum_{x\in\cX,y\in\cY} (P_{\bY|\bx}(y|x) - W(y|x)) I'_W(y|x) \\
                                    &+ O\left(\sum_{x\in\cX,y\in\cY}( P_{\bY|\bx}(y|x) - W(y|x))^2\right),
\end{align}
where $I'_W(y|x) \triangleq\left.\frac{\partial I(\Phi,V)}{\partial V(y|x)}\right|_{V = W}$. 
Specifically, we characterize the first-order and higher-order correction terms of the Taylor
expansion via \Cref{lemma:empirical_mi_main_correction,lemma:empirical_mi_minor_correction}.

\begin{lemma}[First order correction term for mutual information]
    \label{lemma:empirical_mi_main_correction}
    If $\bY \sim \chinn{W}{\bx}$, then
    \begin{equation*}
        \sum_{x\in\cX,y\in\cY} (P_{\bY|\bx}(y|x) - W(y|x)) I'_W(y|x) 
        =
        \sum_x \sum_{j:j\in\cJ_x} Z_{x,j}
        %\frac{1}{N(x|\bx)}
    \end{equation*}
    where $\cJ_x \defeq \SetDef{j}{x_j = x}$, $\Set{Z_{x,j}, x\in\cX, j\in\cJ_x}$ are independent
    random variables,
    and for a given $x$, $\Set{Z_{x,j}, j\in\cJ_x}$ are identically distributed. Furthermore,
    \begin{align*}
        \E{Z_{x,j}} = 0, \quad \forall\, x, j
        \\
        \sum_x \sum_{j:j\in\cJ_x} \Var{Z_{x,j}}
        &= \frac{V(\Phi,W)}{n},
        \\
        \sum_x \sum_{j:j\in\cJ_x} \E{ \abs{Z_{x,j} - \E{Z_{x,j}}}^3  }
        &= \BigO{ \frac{1}{n^2} }.
    \end{align*}
\end{lemma}
\begin{IEEEproof}[Proof of \Cref{lemma:empirical_mi_main_correction}]
Note
\begin{align*}
    \sum_{x\in\cX,y\in\cY} ( P_{\bY|\bx}(y|x) - W(y|x)) I'_W(y|x) &=
    \sum_x \left[\sum_y(P_{\bY|\bx}(y|x) - W(y|x)) I'_W(y|x)\right] \\
    &= \sum_x \left[\sum_{y} P_{\bY|\bx}(y|x) I'_W(y|x) - \sum_{y} W(y|x) I'_W(y|x)\right]\\
    &= \sum_x \left[\frac{1}{N(x|\bx)}\sum_{y} N_{x,y}(\bx,\bY) I'_W(y|x) - E[I'_W(Y|x)]\right]\\
    &= \sum_x \frac{1}{N(x|\bx)}\sum_{j:j\in\cJ_x} \left[I'_W(Y_j|x) - E[I'_W(Y|x)]\right].
\end{align*}
Let $\tlZ_{x,j} \defeq I'_W(Y_j|x) - E[I'_W(Y|x)$ and 
$Z_{x,j} = \frac{1}{N(x|\bx)} \tlZ_{x,j}$, then
$\E{Z_{x,j}} = 0$ and 
\begin{align*}
    \Var{\tlZ_{x,j}} = \Var{I'_W(Y_j|x)} = \Var{I'_W(Y|x)}.
\end{align*}
By straightforward differentiation, 
\begin{equation*}
    I'_W(y|x) 
    = \left.\frac{\partial I(\Phi,V)}{\partial V(y|x)}\right|_{V = W} 
    = \Phi(x)\log\frac{W(y|x)}{\Phi W(y)},
\end{equation*}
thus
\begin{equation*}
    \Var{\tlZ_{x,j}} = \Var{I'_W(Y|x)} = \Phi^2(x) \Var{\log\frac{W(Y|x)}{\Phi W(Y)}}.
\end{equation*}
Therefore
\begin{align*}
    \sum_x \sum_{j:j\in\cJ_x} \Var{Z_{x,j}}
    &=
    \sum_x \sum_{j:j\in\cJ_x} 
    \frac{1}{N(x|\bx)^2} 
    \Var{\tlZ_{x,j}}
    \\
    &=
    \sum_x \frac{1}{n \Pch(x)} 
    \Phi(x)^2 \Var{\log\frac{W(Y|x)}{\Phi W(Y)}} 
    \\
    &= \frac{1}{n} 
    \sum_x \Phi(x) \Var{\log\frac{W(Y|x)}{\Phi W(Y)}} 
    = \frac{V(\Phi,W)}{n}.
\end{align*}
Finally, since any $Z_{x,j}$ is discrete and finite valued variables, the sum of the absolute third
moment of these variables is bounded by some function $r_n = \BigTheta{ \frac{1}{n^2} }$.
\end{IEEEproof}

To investigate the higher order terms, we partition the channel realizations by its closeness to the
true channel distribution $W$. Given input distribution $\Pchn$, we define 
\begin{equation}
    \Xi_n
    \triangleq
    \Xi_n\left( \Pchn \right)
    \triangleq
    \SetDef{V \in \PDSpaceN{\cY|\Pchn}}
    {
    \sum_{x,y}( V(y|x) - W(y|x))^2 
    \leq 
    |\cX|\cdot|\cY|\cdot\frac{\log n}{n}\cdot\frac{1}{\Pchn^{\min}}
    % FIXME: (DW): I removed the 1/n term here. Please verify!
    },
    \label{eq:def_Xi_n}
\end{equation}
where $\Pchn^{\min} \triangleq \min_{x\in\cX}\Pchn(x)$. As shown below in \Cref{lemma:Xi_n}, $\Xi_n$ is
``typical'' in the sense that it contains a channel realization with high probability.
\begin{lemma}
    \label{lemma:Xi_n}
    If $\bx\in\cX^n$ has a type $\Pchn$ and $\bY\in\cY^n$ is the output of the channel $W^n$ with
    input $\bx$, then
    \begin{equation*}
        \Prob{P_{\bY|\bx} \notin \Xi_n } 
        \leq 
        \frac{2|\cX|\cdot|\cY|}{n^2}.
    \end{equation*}
\end{lemma}

\begin{IEEEproof}[Proof of \Cref{lemma:Xi_n}]
 Let $\beta^2 = |\cX|\cdot|\cY|\cdot\frac{\log n}{n}\cdot\frac{1}{\Pchn^{\min}-\frac{1}{n}}$.
\begin{align}
    \Prob{P_{\bY|\bx} \notin \Xi_n }
    &= \Prob{\sum_{a\in\cX,b\in\cY} \left( P_{\bY|\bx}(b|a) - W(b|a)\right)^2 > \beta^2}\nonumber\\
    &\overset{(a)}\leq \Prob{\bigcup_{a\in\cX,b\in\cY} \left\{\left( P_{\bY|\bx}(b|a) - W(b|a)\right)^2 > \tfrac{\beta^2}{|\cX||\cY|}\right\}}\nonumber\\
    &\overset{(b)}\leq \sum_{a\in\cX,b\in\cY} \Prob{\left( P_{\bY|\bx}(b|a) - W(b|a)\right)^2 > \tfrac{\beta^2}{|\cX||\cY|}}\nonumber\\
    &= \sum_{a\in\cX,b\in\cY} \Prob{\left| P_{\bY|\bx}(b|a) - W(b|a)\right| > \tfrac{\beta}{\sqrt{|\cX||\cY|}}},
    \label{eq:bound_Xi_n}
\end{align}
where $(a)$ follows from the fact that in order for a sum of $|\cX||\cY|$ elements to be above $\beta^2$, then at least one of the summands must be above $\beta^2/(|\cX||\cY|)$. $(b)$ follows from the union bound.
For any $a\in\cX,b\in\cY$, we have
\begin{align}
    \quad&\Prob{\left| P_{\bY|\bx}(b|a) - W(b|a)\right| > \tfrac{\beta}{\sqrt{|\cX||\cY|}}}\nonumber\\
    =& \Prob{\left| \tfrac{1}{N_a(\bx)}\sum_{i:x_i = a} \left( \indicator{Y_i=b} - W(b|a) \right) \right| > \tfrac{\beta}{\sqrt{|\cX||\cY|}}}
    \nonumber\\
    \overset{(a)}\leq& 2 \exp\left(-\frac{2 \beta^2 N_a(\bx)}{|\cX|\cdot|\cY| }\right)
    \nonumber\\
    =&2 \exp\left(-\frac{2 \beta^2 n \Pchn(a)}{|\cX|\cdot|\cY| }\right),
    \label{eq:bound_Xi_n_single}
\end{align}
where $(a)$ follows from Hoeffding's inequality (see, e.g. \cite[p. 191]{PollardBook84}). Applying
\Cref{eq:bound_Xi_n_single} to each of the summands of \Cref{eq:bound_Xi_n} gives
\begin{align}
    \Prob{P_{\bY|\bx} \notin \Xi_n }
    &\leq \sum_{a\in\cX,b\in\cY} \Prob{\left| P_{\bY|\bx}(b|a) - W(b|a)\right| > \tfrac{\beta}{\sqrt{|\cX||\cY|}}} \nonumber\\
    &\leq \sum_{a\in\cX,b\in\cY}2 \exp\left(-\frac{2 \beta^2 n \Pchn(a)}{|\cX|\cdot|\cY| }\right)\nonumber\\
    &\leq 2 |\cY|\sum_{a\in\cX} \exp\left(-\frac{2 \beta^2 n \Pchn(a)}{|\cX|\cdot|\cY| }\right)\nonumber\\
    &\leq 2 |\cX|\cdot|\cY| \exp\left(-\frac{2 \beta^2 n \Pchn^{\min}}{|\cX|\cdot|\cY| }\right)\nonumber\\
    &= 2 |\cX|\cdot|\cY| \frac{1}{n^2}.
\end{align}
\end{IEEEproof}

With \Cref{lemma:Xi_n}, we can show that the higher order terms in 
\Cref{eq:empirical_mi_taylor} is in some sense negligible via
\Cref{lemma:empirical_mi_minor_correction}.

\begin{lemma}[Second order correction term for mutual information]
    \label{lemma:empirical_mi_minor_correction}
    If $\bY \sim \chinn{W}{\bx}$, then exists $J = J(\cardinality{\cX}, \cardinality{\cY}, P_\bx)$
    such that
    \begin{align*}
        \Prob{
            \sum_{x\in\cX,y\in\cY}( P_{\bY|\bx}(y|x) - W(y|x))^2
            \geq
            J \flognn
        }
        \leq
        \frac{2 \cardinality{\cX} \cardinality{\cY}}{n^2}
    \end{align*}
\end{lemma}
\begin{IEEEproof}[Proof of \Cref{lemma:empirical_mi_minor_correction}]
Let 
\[
J = |\cX|\cdot|\cY|\cdot\frac{\log n}{n}\cdot\frac{2}{\Pchn^{\min}}
\]
then the lemma follows from the definition of $\Xi_n$ and \Cref{lemma:Xi_n}.
\end{IEEEproof}

Finally, we show the following lemma that is useful for asymptotic analysis.
\begin{lemma}
    \label{lemma:higher_order_terms}
    If $f_n = \BigO{g_n}$, then there exist $\Gamma_n$ and $\Gamma'_n = \BigTheta{\Gamma_n}$ such that
    \begin{align*}
        \Prob{f_n \geq \Gamma'_n}
        &\leq \Prob{g_n \geq \Gamma_n}
        \\
        \Prob{f_n \leq -\Gamma'_n}
        &\leq \Prob{g_n \geq \Gamma_n}
    \end{align*}
    when $n$ sufficiently large.
\end{lemma}
\begin{IEEEproof}[Proof of \Cref{lemma:empirical_mi_minor_correction}]
    By definition there exists $c > 0$ such that when $n$ sufficiently large, 
    \begin{equation*}
        -c g_n \leq f_n \leq c g_n
    \end{equation*}
    Then letting $\Gamma'_n = c \Gamma_n$ completes the proof.
\end{IEEEproof}

\subsection{Proofs for UEP channel coding lemmas}
\label[app]{sec:UEP_proofs}
In this section we provide proofs for \Cref{lemma:rate_redundancy,lemma:UEP_error2}.

\begin{IEEEproof}[Proof for \Cref{lemma:rate_redundancy}]
We directly prove the stronger result where $\Delta R$ is defined according to \Cref{eq:rate_redundancy_eps2}.

By Taylor expansion, we have 
\begin{align*}
    I(\Phi,P_{\bY|\bx}) = I(\Phi,W) &+ \sum_{x\in\cX,y\in\cY} (P_{\bY|\bx}(y|x) - W(y|x)) I'_W(y|x) \\
                                    &+ O\left(\sum_{x\in\cX,y\in\cY}( P_{\bY|\bx}(y|x) - W(y|x))^2\right),
\end{align*}
where $I'_W(y|x) \triangleq\left.\frac{\partial I(\Phi,V)}{\partial V(y|x)}\right|_{V = W}$. 
Let 
$$A(\bY) = \sum_{x\in\cX,y\in\cY} (P_{\bY|\bx}(y|x) - W(y|x)) I'_W(y|x)$$
and
$$B(\bY) = \BigO{\sum_{x\in\cX,y\in\cY}( P_{\bY|\bx}(y|x) - W(y|x))^2},$$ 
then
\begin{align}
    \eps + \delta_n 
    &= 
    \Prob{
    \MIPW{\Pchn}{P_{\bY|\bx}} \leq \MIPW{\Pch}{W} - \Delta R, \bY \sim \chinn{W}{\bx} 
    }
    \nn
    \\
    &= 
    \Prob{
        A(\bY) + B(\bY) \leq - \Delta R, \bY \sim \chinn{W}{\bx} 
    }
    \nn
    \\
    &\overset{(a)}\geq
    \Prob{ A(\bY) + \Gamma_n \leq - \Delta R, \bY \sim \chinn{W}{\bx}  }
    -
    \Prob{ B(\bY) \geq \Gamma_n, \bY \sim \chinn{W}{\bx}  }
    \label{eq:rr_1}
\end{align}
where $\Gamma_n > 0$ and $(a)$ follows from 
\Cref{eq:prob_leq_lower}.
Similarly, \Cref{eq:prob_leq_lower} indicates
\begin{align}
    \eps + \delta_n 
    &= 
    \Prob{
        A(\bY) + B(\bY) \leq - \Delta R, \bY \sim \chinn{W}{\bx} 
    }
    \nn
    \\
    &\leq
    \Prob{ A(\bY) - \Gamma_n \leq - \Delta R, \bY \sim \chinn{W}{\bx}  }
    +
    \Prob{ B(\bY) \leq -\Gamma_n, \bY \sim \chinn{W}{\bx}  }
    \label{eq:rr_2}
\end{align}
Let $\Gamma'_n = J(\Pchn, \cardinality{\cX}, \cardinality{\cY})$ in
\Cref{lemma:empirical_mi_minor_correction}, then from
\Cref{lemma:empirical_mi_minor_correction,lemma:higher_order_terms}, there exists
$\Gamma_n = \BigTheta{\Gamma'_n} = \Oflognn$ such that
\begin{align}
\Prob{ B(\bY) \geq \Gamma_n, \bY \sim \chinn{W}{\bx} } 
\leq \BigO{ \frac{1}{n^2}},
\label{eq:bb1}
\\
\Prob{ B(\bY) \leq -\Gamma_n, \bY \sim \chinn{W}{\bx} }
\leq \BigO{ \frac{1}{n^2}}.
\label{eq:bb2}
\end{align}
In addition, based on \Cref{lemma:empirical_mi_main_correction} and $Q(x) = 1 - Q(-x)$, we can apply
Berry-Esseen theorem (see, e.g., [3, Ch. XVI.5]) and have for any $- \infty < \lambda < \infty$, 
\begin{align}
    \abs{
    \Prob{ A(\bY) \geq \lambda \sigma, \bY \sim \chinn{W}{x} }
        -
        Q\left( \lambda \right)
    }
    \leq
    \frac{T}{\sigma^3},
    \label{eq:clt_approx_geq}
    \\
    \abs{
    \Prob{ A(\bY) \leq -\lambda \sigma, \bY \sim \chinn{W}{x} }
        -
        Q\left( \lambda \right)
    }
    \leq
    \frac{T}{\sigma^3},
    \label{eq:clt_approx_leq}
\end{align}
where $\sigma^2 = V(\Pch, W)/n$ and $T$ is bounded by $c/n^2$, where $c$ is some constant.
Denote $\CondInfoVar{\Pch}{W}$ as $V$, 
apply $\lambda_1 = (\Delta R + \Gamma_n ) / \sigma$ 
and $\lambda_2 = (\Delta R - \Gamma_n) / \sigma$ 
to 
\Cref{eq:clt_approx_geq} 
and
\Cref{eq:clt_approx_leq} 
respectively, 
\begin{align}
    \abs{
    \Prob{ A(\bY) \geq \Delta R + \Gamma_n, \bY \sim \chinn{W}{\bx}  }
    - 
    Q\left( \left( \Delta R + \Gamma_n \right) \sqrt{\frac{n}{{V}}} \right)
    }
    \leq
    \frac{c}{\sqrt{nV^3}},
    \label{eq:sandwich1}
    \\
    \abs{
    \Prob{ A(\bY) \leq - (\Delta R - \Gamma_n), \bY \sim \chinn{W}{\bx}  }
    - 
    Q\left( \left( \Delta R - \Gamma_n \right) \sqrt{\frac{n}{{V}}} \right)
    }
    \leq
    \frac{c}{\sqrt{nV^3}}.
    \label{eq:sandwich2}
\end{align}
Therefore,
\begin{align*}
    Q\left( (\Delta R + \Gamma_n) \sqrt{\frac{n}{{V}}} \right)
    -
    \frac{c}{\sqrt{nV^3}}
    &\overset{\Cref{eq:sandwich1}}\leq
    \Prob{ A(\bY) \geq \Delta R + \Gamma_n, \bY \sim \chinn{W}{\bx}  }
    \\
    &\overset{\Cref{eq:rr_1}}\leq \eps + \delta_n + \Prob{ B(\bY) \geq \Gamma_n, \bY \sim \chinn{W}{\bx} } 
    \\
    &\overset{\Cref{eq:bb1}}= \eps + \Oflognrn.
\end{align*}
Likewise, 
\begin{align*}
    Q\left( (\Delta R - \Gamma_n) \sqrt{\frac{n}{{V}}} \right)
    +
    \frac{c}{\sqrt{nV^3}}
    &\overset{\Cref{eq:sandwich2}}\geq
    \Prob{ A(\bY) \leq - (\Delta R - \Gamma_n), \bY \sim \chinn{W}{\bx}  }
    \\
    &\overset{\Cref{eq:rr_2}}\geq \eps + \delta_n - \Prob{ B(\bY) \leq - \Gamma_n, \bY \sim \chinn{W}{\bx} } 
    \\
    &\overset{\Cref{eq:bb2}}= \eps + \Oflognrn.
\end{align*}
From the smoothness of $Q^{-1}$ around $\eps$,
\begin{align*}
    (\Delta R + \Gamma_n) \sqrt{\frac{n}{{V}}} 
    &\geq
    \Qinv{\eps + \Oflognrn + \frac{c}{\sqrt{nV^3}}}
    =
    \Qinv{\eps} + \Oflognrn,
    \\
    (\Delta R - \Gamma_n) \sqrt{\frac{n}{{V}}} 
    &\leq
    \Qinv{\eps + \Oflognrn - \frac{c}{\sqrt{nV^3}}}
    =
    \Qinv{\eps} + \Oflognrn.
\end{align*}
Therefore,
\begin{align*}
    \Delta R
    &\geq
    \sqrt{\frac{V}{{n}}} \Qinv{\eps} + 
    \sqrt{\frac{V}{{n}}} \Oflognrn - \Gamma_n
    =
    \sqrt{\frac{V}{{n}}} \Qinv{\eps} + \Oflognn,
    \\
    \Delta R
    &\leq
    \sqrt{\frac{V}{{n}}} \Qinv{\eps} + 
    \sqrt{\frac{V}{{n}}} \Oflognrn + \Gamma_n
    =
    \sqrt{\frac{V}{{n}}} \Qinv{\eps} + \Oflognn,
\end{align*}
and finally
\begin{equation*}
    \Delta R
    =
    \sqrt{\frac{V}{{n}}} \Qinv{\eps} + \Oflognn.
\end{equation*}
\end{IEEEproof}

Before proving \Cref{lemma:UEP_error2}, we include the following
lemma~\cite{CsiszarLosslessJointExponent} for completeness.
\begin{lemma}[\mbox{\cite[Lemma 6]{CsiszarLosslessJointExponent}}]
    Given $\cX$ and positive integers $n$, $k_n$, let
    \[
    \eta_n \defeq \frac{2}{n}\left( 
        \cardinality{\cX}^2 + \log(n+1) + \log k_n + 1
    \right).
    \]
    Then for arbitrary (not necessarily distinct) distributions $\Pch_i \in \PDSpaceXN$
    and positive integers $N_i$ with 
    \begin{equation*}
        \navelog N_i \leq \Entropy{\Pch_i} - \eta_n, 
        \quad
        i = 1, 2, \ldots, m,
        %\label{}
    \end{equation*}
    there exist $m$ disjoint sets $\cA_i \subset \cX^n$ such that
    \begin{equation*}
        \cA_i \subset \TypeSetn{\Pch_i}, 
        \cardinality{\cA_i} = N_i,
        \quad
        i = 1, 2, \ldots, m,
        %\label{}
    \end{equation*}
    and
    \begin{equation*}
        \cardinality{ \TypeShell{\brV}{\bx} } 
        \leq
        N_j \exp\left\{ -n \left[ \MIPW{\Pch_i}{\brV} - \eta_n \right] \right\}
        \text{ if }
        \bx \in \cA_i
        %\label{}
    \end{equation*}
    for every $i, j$ and $\brV: \cX^n \rightarrow \cX^n$, except for the case $i=j$ and $\brV$ is
    the identity matrix.
    \label{lemma:packing}
\end{lemma}

\begin{IEEEproof}[Proof for \Cref{lemma:UEP_error2}]
For $\bx' \in \cA_j$, $\bx' \neq \bx$, 
let the joint type for the triple 
$(\bx, \bx', \by)$ be given as the joint distribution of
%has a fixed joint type. Let this joint type 
RV's $X, X', Y$.
Then from
\Cref{lemma:packing}, we can find $\Set{\cA_i}$ such that
$\cA_i \subset \TypeSetn{\Pch_i}$ and
$\navelog N_i \leq \Entropy{\Pch_i} - \eta_n$, 
thus $X$ has distribution $\Pch_i$ and $X'$ has distribution $\Pch_j$. 
In addition, define
\begin{equation*}
    \cB_{V}
    \defeq
    \cB_{V}(\bx)
    \defeq
    \SetDef{\by \in \TypeShelln{V}{\bx}}{
    %(\bx, \bx', \by) \text{ has a joint type } P_{X, X', Y}
    %\text{ and }
    \exists \, \bx' \neq \bx
    \text{ such that }
    %\MIxy{\bx'}{\by} \geq R_i + \gamma
    \bx' \in \cA_j 
    \text{ and }
    \MIxy{\bx'}{\by} - R_j \geq \gamma
    },
    %\label{}
\end{equation*}
then the cardinality of $\cup_V \cB_{V}$ is upper bounded by
%we can find sets $\cA_i$ such that
\begin{align*}
\cardinality{\cup_V \cB_{V}}
&\leq 
N_j 
\exp\left\{ -n 
    \left[ \MIXY{X,X'}{Y} - H(Y|X) - \eta_n \right]
\right\}
\\
&\leq 
N_j 
\exp\left\{ n H(Y|X)
-n \posfunc{\MIXY{X,X'}{Y} - \eta_n}
\right\}
\end{align*}

Then for $\by \in \cB_{V}$,
\begin{align*}
    \chio{W}{\by}{\bx} 
    &= \exp\left\{ -n\left[ \CKLD{V}{W}{\Pch_i} + H(V|\Pch_i) \right] \right\}
\end{align*}

Note that $\MIxy{\bx'}{\by} -R_j \geq \gamma$ implies $\MIXY{X'}{Y} - R_j \geq \gamma$, 
and $\MIXY{X,X'}{Y} \geq \MIXY{X}{Y}$, 
\[
\MIXY{X,X'}{Y} - R_j \geq
\MIXY{X'}{Y} - R_j \geq \gamma
\]

Hence, 
\begin{align*}
    \chion{W}{\cB_{V}}{\bx} 
&\leq
    N_j 
    \exp\left\{ n H(Y|X)
    -n \posfunc{\MIXY{X,X'}{Y} - \eta_n}
    \right\}
    \exp\left\{ -n\left[ \CKLD{V}{W}{\Pch_i} + H(V|\Pch_i) \right] \right\}
\\
&= N_j \exp\left\{ -n\left[ \CKLD{V}{W}{\Pch_i} + \posfunc{\MIXY{X,X'}{Y} - \eta_n} \right] \right\}
\\
&\leq
N_j \exp\left\{ -n\left[ \CKLD{V}{W}{\Pch_i} + \posfunc{R_j + \gamma - \eta_n} \right] \right\}
\end{align*}

And 
\begin{align*}
    \Prob{\MIxy{\bx'}{\by} - R_j \geq \gamma}
    &\leq \chion{W}{\bigcup_{V}\cB_{V}}{\bx}
    \\
    &\leq (n+1)^{\cardinality{\cX}^2\cardinality{\cY}}
    N_j \exp\left\{ -n\left[ \posfunc{R_j + \gamma - \eta_n} \right] \right\}
    %\\
    %&\leq m_n (n+1)^{\cardinality{\cX}^2\cardinality{\cY}}
    %\exp\left\{ -n\left[ \posfunc{\gamma - \eta_n}  \right] \right\}
    %\quad
    %\text{when } \gamma \geq \eta_n
\end{align*}
\end{IEEEproof}

\section{Proofs for JSCC dispersion}
This appendix contains proofs for results in \Cref{sec:JSCC}.  Similar to the development in
\Cref{sec:proofs_UEP}, we start by analyzing the Taylor expansion of the distortion-rate function
in \Cref{sec:DR_taylor}, then prove the relevant key lemmas \Cref{sec:JSCC_proofs}.

\subsection{Analysis of the distortion-rate function }
\label[app]{sec:DR_taylor}
In this section, we investigate that Taylor expansion of $R(P_\bS,D_n)$.
Denote the partial derivatives of $D(P,R)$ at $R = I(\Phi,W)$ and $Q=P$ as
\begin{align*}
D'_R 
&\triangleq \left.\frac{\partial D(P,R)}{\partial R}\right|_{R = I(\Phi,W)},
\\
D'_P(s) 
&\triangleq \left.\frac{\partial D(Q,R)}{\partial Q(s)}\right|_{Q = P}.
\end{align*}
Assuming $D(\cdot,\cdot)$ is smooth, 
Taylor expansion gives
\begin{align}
    D(P_\bS, \rho I(\Phi,P_{\bY|\bx})+\xi'_n) 
    &= D(P,\rho I(\Phi,W)) 
    \nn
    \\
    &+ \sum_{s=1}^{|\cS|}(P_\bS(s) - P(s)) D'_P(s) 
    \nn
    \\
    &+ (\rho I(\Phi,P_{\bY|\bx}) +\xi_n'- \rho I(\Phi,W)) D'_R 
    \nn
    \\
    &+ O\left(\sum_{s=1}^{|\cS|}(P_\bs(s) - P(s))^2 + (\rho I(\Phi,P_{\bY|\bx}) +\xi_n'- \rho I(\Phi,W))^2\right)
    \nn
    \\
    &= D(P,\rho I(\Phi,W)) 
    \nn
    \\
    &+ \sum_{s=1}^{|\cS|} \left(P_\bS(s) - P(s)\right) D'_P(s) 
    + \rho D'_R \sum_{x,y} \left( P_{\bY|\bx}(y|x) - W(y|x) \right) I'_W(y|x)
    \\
    &+ B(\bS, \bY, \xi'_n),
\end{align}
where $\xi'_n = \BigO{\log n / n}$, and
the correction term is 
\begin{align}
    B(\bS, \bY, \xi'_n) 
    \defeq& \xi_n'D'_R +O\left(\sum_{x,y}( P_{\bY|\bx}(y|x)-W(y|x))^2\right)\nonumber\\
     &+ O\left(\sum_{s=1}^{|\cS|}(P_\bs(s) - P(s))^2 + (\rho I(\Phi,P_{\bY|\bx}) +\xi_n'- \rho I(\Phi,W))^2\right). 
    \label{eq:D(R)_correction_terms}
\end{align}
For notational simplicity, we define
\begin{equation}
    A(\bS,\bY)
    \defeq
    \sum_{s=1}^{|\cS|} \left(P_\bS(s) - P(s)\right) D'_P(s) 
    + \rho D'_R \sum_{x,y} \left( P_{\bY|\bx}(y|x) - W(y|x) \right) I'_W(y|x)
    \label{eq:sum_main_correction}
\end{equation}

The lemmas in this subsection is organized as follows.  \Cref{lemma:1st_order_src} shows that the
first order terms of the Taylor expansion of $R(P_\bS,D_n)$ with respect to $P$ can be represented
as the sum of $n$ \iid random variables. Then \Cref{lemma:1st_order_DR} shows that $A(\bS,\bY)$ can
be represented represented as the sum of $n+m$ \iid random variables. Finally,
\Cref{lemma:Xi_n_property,lemma:quadratic_terms} together with
\Cref{lemma:Xi_n,lemma:higher_order_terms} shows that the higher order terms in the Taylor expansion
is negligible, as summarized in \Cref{lemma:2nd_order_DR}.

\begin{lemma}
    \label{lemma:1st_order_src}
    Under the conditions of \Cref{lemma:JSCCRedundancy},
    \begin{equation*}
    \sum_{s\in\cS}(P_\bS(s) - P(s)) D'_P(s) 
    = \sum_{i=1}^n \tlZ_i
    \end{equation*}
    where $\{\tlZ_i, i=1,2,\cdots,n\}$ are \iid random variables such that
    \begin{align*}
        \E{\tlZ_i} &= 0
        \\
        \Var{\tlZ_i} &= \frac{V_D}{n^2}
    \end{align*}
    where $V_D = V_S \cdot (D'_R)^2$. 
\end{lemma}
\begin{IEEEproof}
\begin{align*}
    \sum_{s\in\cS}(P_\bS(s) - P(s)) D'_P(s) 
    &= \frac{1}{n}\sum_{i=1}^n D'_P(S_i) -  \sum_{s\in\cS} P(s) D'_P(s)\\
    &= \frac{1}{n}\sum_{i=1}^n D'_P(S_i) -  E[ D'_P(S)]\\
    %&= \frac{1}{n}\sum_{i=1}^n \left[D'_P(S_i) -  E[ D'_P(S)]\right],
\end{align*}
Let $\tlZ_i \defeq D'_P(S_i) -  E[ D'_P(S)]$, then
\begin{align*}
    \E{\tlZ_i} &= 0,
\end{align*}
and
\begin{align*}
  \Var{D'_P(S_i) -  E[ D'_P(S)]} 
  &= \Var{D'_P(S)}.
\end{align*}
By elementary calculus it can be shown that for all $s \in \cS$,
\begin{equation*}
    D'_P(s) 
    = \frac{\partial D(P,R)}{\partial P(s)} 
    = - \frac{\partial R(P,D)}{\partial P(s)} \frac{\partial D(P,R)}{\partial R} 
    = - R'(s) D'_R.
\end{equation*}
Therefore, 
\begin{equation*}
    V_D = \Var{D'_P(S)} = \Var{R'(S)}(D'_R)^2 = V_S \cdot (D'_R)^2.
\end{equation*}
\end{IEEEproof}

\begin{lemma}[First order correction term for distortion-rate function]
    \label{lemma:1st_order_DR}
    Under the conditions of \Cref{lemma:JSCCRedundancy},
    \Cref{eq:sum_main_correction}, \ie, $A(\bS,\bY)$
    is the sum of $n+m$ independent random variables, whose sum of variance is 
    \begin{equation*}
        \frac{1}{n}
        \left[ 
        \rho(D'_R)^2 V_S
        + \rho(D'_R)^2 \CondInfoVar{\Pch}{W}
        + \BigO{ \frac{\log n}{n}}
        \right]
        \label{eq:V_asymptotic}
    \end{equation*}
    and sum of the absolute third moment is bounded by some constant.
\end{lemma}
\begin{IEEEproof}[Proof for \Cref{lemma:1st_order_DR}]
According to \Cref{lemma:1st_order_src,lemma:empirical_mi_main_correction},
\Cref{eq:sum_main_correction} 
can be interpreted as the sum of $n+m$ independent random variables.
Let $\sigma_n^2$ be the sum of the variance of these $n+m$ variables, then
\begin{align}
    \sigma^2_n 
    &=  n \frac{1}{n^2}V_D + \sum_{x\in\cX} m \Phi_m(x) \left(\frac{\rho D'_R}{m \Phi_m(x)}\right)^2V_C(x) 
    \nn
    \\
    &= \frac{1}{n}V_D + \sum_{x\in\cX} \frac{(\rho D'_R)^2}{m \Phi_m(x)} V_C(x).
    \nn
    \\
    &= \frac{1}{n}
    \left[ 
    V_D + \rho(D'_R)^2 \CondInfoVar{\Pch_m}{W}
    \right]
    \nn
    \\
    &= \frac{1}{n}
    \left[ 
    \rho(D'_R)^2 V_S
    + \rho(D'_R)^2 \CondInfoVar{\Pch}{W}
    + \BigO{ \frac{\log n}{n}}
    \right]
\end{align}
Define $r$ to be the sum of the absolute third moment of these variables. Since these are discrete
and finite valued variables, $r$ is bounded by $\frac{1}{n^2}J_3$, for some constant $J_3$.

\end{IEEEproof}

To investigate the higher order terms, we partition the source type by its closeness to the
source distribution $P$. Given source distribution $P$, we define 
\begin{equation}
   \Omega_n \defeq \Omega_n(P) \defeq 
   \left\{Q \in \Types_n: \|P-Q\|_2^2 \leq |\cS| \frac{\log n}{n} \right\}. %\nonumber \\
\end{equation}
In addition, we show the following property of set $\Xi_n$ (defined in \Cref{eq:def_Xi_n} in 
\Cref{sec:proofs_empirical_MI}):
\begin{lemma}
    \label{lemma:Xi_n_property}
    If $P_{\by|\bx} \in \Xi_n$, then
    \begin{equation}
        \left(I(\Phi,P_{\by|\bx}) - I(\Phi,W)\right)^2 = O\left(\frac{\log n}{n}\right).
    \end{equation}
\end{lemma}
\begin{IEEEproof}
    By definition of $\Xi_n$,
    \begin{equation}
        \sum_{x,y}( P_{\by|\bx}(y|x) - W(y|x))^2 =  O\left(\frac{\log n}{n}\right),
    \end{equation}
    and therefore
    \begin{equation}
        \max_{x,y}( P_{\by|\bx}(y|x) - W(y|x))^2 = O\left(\frac{\log n}{n}\right).\label{eqn:bound_max_diff_cond_type}
    \end{equation}
    The zero-th order Taylor approximation of $I(\Phi,P_{\by|\bx})$ around $W=P_{\by|\bx}$ is given by
    \begin{align}
        I(\Phi,P_{\by|\bx})
        &= I(\Phi,W) + O\left(\sum_{x,y} \Bigl| P_{\by|\bx}(y|x)-W(y|x)\Bigr| \right)\\
        &= I(\Phi,W) + O\left(\max_{x,y} \Bigl| P_{\by|\bx}(y|x)-W(y|x)\Bigr| \right),
    \end{align}
    therefore
    \begin{align}
        (I(\Phi,P_{\by|\bx}) - I(\Phi,W))^2 = O\left(\max_{x,y} \Bigl| P_{\by|\bx}(y|x)-W(y|x)\Bigr|^2 \right),
    \end{align}
    and the required result follows from \Cref{eqn:bound_max_diff_cond_type}.
\end{IEEEproof}

The bounding of $B(\bS, \bY, \xi'_n)$ is mainly based on the following lemma.

\begin{lemma}
    \label{lemma:quadratic_terms}
    There exists constant $J > 0$ such that
    \begin{align*}
        \Prob{
        \sum_{x,y}( P_{\bY|\bx}(y|x)-W(y|x))^2
        + \sum_{s=1}^{|\cS|}(P_\bs(s) - P(s))^2 
        + (\rho I(\Phi,P_{\bY|\bx}) +\xi_n'- \rho I(\Phi,W))^2
        \geq
        J \frac{\log n}{n}
        }
        \\
        \leq \BigO{\frac{1}{n^2}}
    \end{align*}
\end{lemma}
\begin{IEEEproof}
Based on \Cref{lemma:Xi_n_property}, we have
\begin{align*}
    &\Prob{
    \sum_{x,y}( P_{\bY|\bx}(y|x)-W(y|x))^2
    + \sum_{s=1}^{|\cS|}(P_\bs(s) - P(s))^2 
    + (\rho I(\Phi,P_{\bY|\bx}) +\xi_n'- \rho I(\Phi,W))^2
    \geq
    J \frac{\log n}{n}
    }
    \\
    &\leq \Prob{P_\bS \notin \Omega_n \text{ or }  P_{\bY|\bx} \notin \Xi_n}\\
    &\leq \Prob{P_\bS \notin \Omega_n} + \Prob{P_{\bY|\bx} \notin \Xi_n}\\
    &\overset{(a)}\leq \frac{2|\cS|}{n^2} + \frac{2|\cX|\cdot|\cY|}{m^2}\\
    &=O\left(\frac{1}{n^2}\right).
\end{align*}
(a) follows from \Cref{lemma:Xi_n} and \cite[Lemma 2]{AmirYuvalDCC}.
\end{IEEEproof}

\begin{lemma}[Second order correction term for distortion-rate function]
    \label{lemma:2nd_order_DR}
    For $\xi'_n = \Oflognn$, there exists 
    $\Gamma_{n,1} = \Oflognn$ 
    and
    $\Gamma_{n,2} = \Oflognn$ 
    such that
    \begin{align}
        \Prob{ 
            B(\bS, \bY, \xi'_n)
            > 
            \Gamma_{n,1}
            }
        \leq
        \BigO{\frac{1}{n^2}}
        \label{eq:B_lb}
        \\
        \Prob{ 
            B(\bS, \bY, \xi'_n)
            < 
            -\Gamma_{n,2}
            }
        \leq
        \BigO{\frac{1}{n^2}}
    \end{align}
\end{lemma}
\begin{IEEEproof}
    Let $\Gamma_{n,1} = \xi'_n D'_R + (J + \abs{D'_R}) \log n / n$
    and
    $\Gamma_{n,2} = - \xi'_n D'_R + (J + \abs{D'_R}) \log n / n$, 
    where the $J$ is given by \Cref{lemma:quadratic_terms}, 
    then the proof follows from 
    \Cref{lemma:quadratic_terms}
    and
    \Cref{lemma:higher_order_terms}.
\end{IEEEproof}

\subsection{Proofs for JSCC lemmas}
\label[app]{sec:JSCC_proofs}
%\label{sec:JSCC_achievability_proofs}
This section first shows \Cref{lemma:JSCCRedundancy} (JSCC Distortion Redundancy
Lemma), upon which proofs for both the achievability and converse of the main theorem builds. Then
it shows the proof for \Cref{lemma:jscc_converse_given_types}, which is essential for establishing
the converse result.

\begin{IEEEproof}[Proof for \Cref{lemma:JSCCRedundancy}]
We directly prove the stronger result where $D_n$ is defined according to \Cref{eq:def_DeltaDn_pert}.

We first note that for $D_n$,
\begin{equation}
    \label{eq:DeltaDn_property}
    \Prob{R(P_\bS, D_n) \geq \rho I(\Phi_m,P_{\bY|\bx}) + \xi_n} \geq \eps + \zeta_n.
\end{equation}

By \Cref{lemma:mutual_info_cont}, for any conditional type $V$, there is a constant 
$J_1 = J_1(|\cX|,|\cY|)$ such that
\begin{equation*}
    |I(\Phi_m,V) - I(\Phi,V)| \leq J_1 \frac{\log m}{m},
\end{equation*}
Therefore, 
\begin{align}
    \eps + \zeta_n
    &\leq \Prob{R(P_\bS,D_n) \geq \rho I(\Phi_m,P_{\bY|\bx}) + \xi_n}
    \nn
    \\
    &\leq \Prob{R(P_\bS,D_n) \geq \rho I(\Phi,P_{\bY|\bx}) - J_1 \tfrac{\log m}{m} + \xi_n}
    \nn
    \\
    &= \Prob{R(P_\bS, D_n) \geq \rho I(\Phi,P_{\bY|\bx}) +\xi'_n}
    \nn
    \\
    &= \Prob{D_n \leq D\left(P_\bS, \rho I(\Phi,P_{\bY|\bx}) +\xi'_n\right)},
    \label{eq:distortion_dispersion_1}
\end{align}
where $xi'_n = \BigO{\log n/n}$.
Let $\Delta D_n \defeq D_n - D^*$, \Cref{eq:distortion_dispersion_1} now becomes
\begin{align*}
    \eps + \zeta_n
    &= \Prob{D_n \leq D(P_\bS, \rho I(\Phi,P_{\bY|\bx}) +\xi'_n)} \\
    &= \Prob{\Delta D_n \leq A(\bS,\bY) + B(\bS, \bY, \xi'_n)}.
\end{align*}
Applying \Cref{eq:prob_geq_lower,eq:prob_geq_upper}
gives
\begin{align*}
    \eps + \zeta_n
    \leq& 
    \Prob{ 
        A(\bS,\bY) + \Gamma_{n,1}
        \geq 
        \Delta D_n} 
        + \Prob{B(\bS, \bY, \xi'_n) > \Gamma_{n,1}}
    \\
    \eps + \zeta_n
    \geq& 
    \Prob{ 
        A(\bS,\bY) - \Gamma_{n,2}
        \geq 
        \Delta D_n} 
    - \Prob{B(\bS, \bY, \xi'_n) < - \Gamma_{n,2}}
\end{align*}
From \Cref{lemma:2nd_order_DR,lemma:higher_order_terms} we have
\begin{align*}
    \Prob{B(\bS, \bY, \xi'_n) < - \Gamma_{n,2}}
    &\leq \BigO{\frac{1}{n^2}}
    \\
    \Prob{B(\bS, \bY, \xi'_n) > \Gamma_{n,1}}
    &\leq \BigO{\frac{1}{n^2}}
\end{align*}
Since $\zeta_n = O\left(\frac{\log n}{\sqrt n}\right)$, we absorb the $\BigO{1/n^2}$ terms and have:
\begin{align*}
    \eps + O\left(\frac{\log n}{\sqrt n}\right)
    &\geq
    \Prob{ A(\bS,\bY) \geq \Delta D_n - \Gamma_{n,1}}
    \\
    \eps + O\left(\frac{\log n}{\sqrt n}\right)
    &\leq 
    \Prob{ A(\bS,\bY) \geq \Delta D_n + \Gamma_{n,2} },
\end{align*}
Based on \Cref{lemma:1st_order_DR},
by the (non-i.i.d. version of the) Berry-Esseen theorem (\cite[XVI.5, Theorem 2]{Feller1971}) we have
that for any $a$ and $n$, 
\begin{align*}
&\abs{
    \Prob{
        A(\bS,\bY)
        \geq 
        \lambda \cdot \sigma_n 
    } - Q(\lambda)
}
\leq \frac{6T_n}{\sigma_n^3} =O\left(\frac{1}{\sqrt n}\right),
\end{align*}
where $T_n$ is bounded by $c/n^2$, with $c$ being a constant.
Let 
$\lambda_1=(\Delta D_n - \Gamma_{n,1})/\sigma$
and
$\lambda_2=(\Delta D_n + \Gamma_{n,2})/\sigma$,
then,
\begin{align*}
    \eps + O\left(\frac{\log n}{\sqrt n}\right)
    \geq 
    Q((\Delta D_n - \Gamma_{n,1})/\sigma) + O\left(\frac{1}{\sqrt n}\right),
    \\
    \eps + O\left(\frac{\log n}{\sqrt n}\right)
    \leq 
    Q((\Delta D_n + \Gamma_{n,2})/\sigma) + O\left(\frac{1}{\sqrt n}\right),
\end{align*}
absorbing the $O\left(\frac{1}{\sqrt n}\right)$ on the right hand side, we have
\begin{align*}
    \eps + O\left(\frac{\log n}{\sqrt n}\right)
    \geq Q((\Delta D_n - \Gamma_{n,1})/\sigma_n),
    \\
    \eps + O\left(\frac{\log n}{\sqrt n}\right)
    \leq Q((\Delta D_n + \Gamma_{n,2})/\sigma_n).
\end{align*}
From the smoothness of $Q^{-1}$ around $\eps$, 
noting $\Gamma_{n,i} = \BigO{\log n / n}, i = 1, 2$ and 
replace $\sigma_n$ as in \Cref{eq:V_asymptotic}, we obtain
\begin{align}
    \label{eq:Dn_JSCC_dispersion_Dn}
    \Delta D_n 
    \leq D'_R 
    \sqrt\frac{V_C + \rho V(\Phi,W) }{n}Q^{-1}(\eps) + \BigO{\frac{\log n}{n}},
    \\
    \Delta D_n 
    \geq D'_R 
    \sqrt\frac{V_C + \rho V(\Phi,W) }{n}Q^{-1}(\eps) + \BigO{\frac{\log n}{n}}.
\end{align}
Therefore, 
\begin{align}
    \Delta D_n 
    = 
    D'_R \sqrt\frac{V_C + \rho V(\Phi,W) }{n}Q^{-1}(\eps) + \BigO{\frac{\log n}{n}}.
\end{align}
We add $D^*$ and apply $R(P,D)$ to both sides of \Cref{eq:Dn_JSCC_dispersion_Dn}. With the Taylor approximation we have
\begin{equation*}
    R(P, D_n)  
    =
    I(\Phi,W) + \sqrt\frac{V_S + \rho V(\Phi,W) }{n}Q^{-1}(\eps)|D'_R|R'_D 
    + \BigO{\frac{\log n}{n}}.
\end{equation*}
where $R'_D \triangleq \frac{\partial R(P,D)}{\partial D}$. Finally, note that $D'_R$ is negative, and combined with the fact that $D'_R R'_D = 1$ we have the required
\begin{equation*}
    R(P, D_n)  
    =
    I(\Phi,W) - \sqrt\frac{V_S + \rho V(\Phi,W) }{n}\Qinv{\eps}
    + \BigO{\frac{\log n}{n}}.
\end{equation*}

In order to establish \Cref{eq:DistortionRedundancyJSCC_unrestricted}, write:
\begin{equation*}   \eps_n \triangleq \Prob{R(P_\bS,D_n) > \rho I(\Phi_m(\bS),P_{\bY|\bx}) + \xi_n} = \sum_{\bs \in \cS^n}  \Prob{\bS=\bs}  \Prob{  I(\Phi_m(\bs),P_{\bY|\bx}) < T_n(P_\bs) }, \end{equation*} where
\[T_n(P_\bs) \triangleq \frac{R(P_\bs,D_n) - \xi_n}{\rho}. \] 
Clearly, the optimal $\Phi_m(\bs)$ is only a function of $T_n(P_\bs)$. Thus,
\begin{equation} \label{eq:sum_for_end_lemma} \eps_n \geq \sum_t \Prob{T_n(P_\bs) = t} \Prob{
    I(\Phi_m(t),P_{\bY|\bx}) < t }. 
\end{equation} 
Without loss of generality we restrict the thresholds to those satisfying 
\begin{equation} \label{eq:interesting_thresholds} t \geq C(W) -
\BigO{\frac{\log n}{n}}, 
\end{equation} 
since otherwise the theorem is satisfied trivially.
Now define the set 
\[\Pi(W, \delta) \triangleq \{\Phi \in \PDSpaceX : \exists
\Phi^*\in\Pi(W) :  \|\Phi - \Phi^*\| \leq \delta \} . \] Since $I(\Phi,W)$ is concave in
$\Phi$, it follows that \[\sup_{\Phi\notin \Pi(W,\delta)} I(\Phi,W) = C(W) -
\epsilon(\delta) \] where $\epsilon(\delta)>0$ for any $\delta>0$. Thus, for thresholds that
satisfy \Cref{eq:interesting_thresholds} and for $\Phi\notin \Pi(W,\delta)$ (for any choice
of $\delta>0$): \[  \lim_{n\rightarrow\infty} \Prob{  I(\Phi,P_{\bY|\bx}) < t } = 1. \] It
follows that we may restrict $\Phi_m(t)$ in \Cref{eq:sum_for_end_lemma} to any set
$\Pi(W,\delta)$ with $\delta>0$. Since inside that set the Hessian of $I(P,W)$ (as a
function of $W$) can be uniformly bounded (see \cite[Appendix I]{PolyanskiyPVFiniteLength10}), we
have that \Cref{eq:sandwich1,eq:sandwich2} holds uniformly (i.e. with the same constant $A$) for all
$\Phi\in \Pi(W,\delta)$. 
Consequently, 
\[  
\Prob{  I(\Phi_m(t),P_{\bY|\bx}) < t } \geq 1 - Q\left(\Bigl(t-I(\Phi_m(t),W)\Bigr) \sqrt{\frac{n}{V(\Phi_m(t),W)}}  \right) + \BigO{\frac{1}{\sqrt{n}}} 
\] 

Since without the last correction term the probability is minimized by any $\Phi^*(W) \in \Pi(W)$
and that correction term is uniform, we have that 
\[  
\Prob{  I(\Phi_m(t),P_{\bY|\bx}) < t } \geq  \Prob{  I(\Phi^*(W) ,P_{\bY|\bx}) < t } - \BigO{\frac{1}{\sqrt{n}}}. 
\] 
Then, \Cref{eq:sum_for_end_lemma} becomes:
\[ \eps_n  + \BigO{\frac{1}{\sqrt{n}}} \geq  \sum_t \Prob{T_n(P_\bs) = t} \Prob{  I(\Phi^*(W) , P_{\bY|\bx}) < t } =  \Prob{R(P_\bS,D_n) > \rho I(\Phi^*(W),P_{\bY|\bx}) + \xi_n}  \] 
Since the $\BigO{\nicefrac{1}{\sqrt{n}}}$ term may be included in a $\xi_n$ sequence, it follows
that one cannot do better, to the approximation required, then using a fixed input type $\Phi^*(W)$
for all source strings, resulting in  \Cref{eq:DistortionRedundancyJSCC_unrestricted}. 

\end{IEEEproof}

%\label{sec:JSCC_converse_proofs}
To show the converse of the JSCC problem define in \Cref{sec:intro}, we first upper bound the
fraction of source codeword that is $D$-covered by a given reconstruction sequence.

\begin{lemma}[Restricted $D$-ball size]
    \label{lemma:d-ball-size}
    Given source type $P$ and a reconstruction sequence $\hat{\bs}$, define restricted $D$-ball as
    \begin{equation*}
        B(\hat{\bs}, P, D)
        \defeq 
        \SetDef{\bs \in \TypeSetn{P}}{ d(\bs, \hat{\bs}) \leq D}.
        %\label{eq:d-ball}
    \end{equation*}
    Then
    \begin{align*}
        \cardinality{
        B(\hat{\bs}, P, D)
        }
        \leq
        (n+1)^{\cardinality{\cS}{\cardinality{\hat{\cS}}}}
        \exp \left\{ n \left[ H(P) - R(P, D) \right] \right\}
    \end{align*}
\end{lemma}

\begin{IEEEproof}
    Let $P\in\PDSpaceN{\cS}$ be a given type and let $Q$ be the type of $\hat{\bs}$.
    %Let $\P \times \Lambda \in \PDSpaceN{\cS \times \hat{\cS}}$ be the joint type such that
    Then the size of the set of source codewords with type $P$ that are $D$-covered by $\hat{\bs}$ is
    \begin{align*}
        \cardinality{ B(\hat{\bs}, P, D) }
        &= 
        \cardinality{
            \bigcup_{\Lambda: \substack{
            \E{d(S, \hS)} \leq D,
            \\
            P \Lambda = Q
            }} 
                \SetDef{\bs \in \TypeSetn{P}}
                {P_{\bs, \hat{\bs}} = P \times \Lambda}
            }
    \end{align*}
    Note there are at most $(n+1)^{\cardinality{\cS}{\cardinality{\hat{\cS}}}}$ joint types, and
    \begin{align*}
            \SetDef{\bs \in \TypeSetn{P}}
            {P_{\bs, \hat{\bs}} = P \times \Lambda}
            =
            \TypeShelln{\tlLambda}{\hat{\bs}},
    \end{align*}
    where $\tlLambda$ is the reverse channel from $\hat{\cS}$ to $\cS$ such that
    $Q \times \tlLambda = P \times \Lambda$.
    Therefore, 
    \begin{align*}
        \cardinality{ B(\hat{\bs}, P, D) }
        &\leq
        \sum_{
            \tlLambda: \substack{
            \Ep{Q,\tlLambda}{d(\hS, S)} \leq D,
            }} 
            \cardinality{ \TypeShelln{\tlLambda}{\hat{\bs}} }
        \\
        &\leq
        (n+1)^{\cardinality{\cS}{\cardinality{\hat{\cS}}}}
            \exp\left[ n 
                \max_{
                    \tlLambda: \substack{
                    \Ep{Q,\tlLambda}{d(\hS, S)} \leq D,
                    }}
                \CondEntropy{\tlLambda}{Q} 
                \right]
    \end{align*}
    Note 
    \begin{align*}
        R(P,D) 
        &= \min_{\Lambda: \Ep{P,\Lambda}{d(S,\hat{\cS})} \leq D}\MIPW{P}{\Lambda}
        \\
        &= H(P) - \max_{\tlLambda: \Ep{Q,\tlLambda}{d(\hS,S)} \leq D} \CondEntropy{\tlLambda}{Q},
    \end{align*}
    hence
    \begin{equation*}
        \cardinality{ B(\hat{\bs}, P, D) }
        \leq
        (n+1)^{\cardinality{\cS}{\cardinality{\hat{\cS}}}}
            \exp \left\{ n \left[ H(P) - R(P, D) \right] \right\}
    \end{equation*}
\end{IEEEproof}

\begin{remark}
    Lemma 3 in~\cite{ZhangYangWei97}, is similar to \Cref{lemma:d-ball-size}. However, it does not
    bound the size of the restricted $D$-ball uniformly, and we choose to prove
    \Cref{lemma:d-ball-size}, which is necessary for proving  
    \Cref{lemma:jscc_converse_given_types}.
\end{remark}

\begin{IEEEproof}[Proof for \Cref{lemma:jscc_converse_given_types}]
    In our proof, we first bound the denominator in \Cref{eq:converse_given_types} uniformly
    for all $\bs_i$, and then bound the sum of the numerator over all $\bs_i$, as done in
    \cite{DueckKorner} for the channel error exponent.

    \paragraph{Bounding the denominator}
    Based on standard results in method of types~\cite{Csiszar81}, 
    for $f(\bs) \in \TypeSetn{\Pch}$, 
    \begin{align*}
        (m+1)^{-\cardinality{\cX}\cardinality{\cY}}
        \exp\left\{ m \CondEntropy{V}{\Pch} \right\}
        \leq
        \cardinality{\TypeShelln[m]{V}{f(\bs)}}
        %\leq
        %\exp\left\{ m \CondEntropy{V}{\Pch} \right\}
    \end{align*}
    Hence
    \begin{align*}
        \frac{1}{ \cardinality{\TypeShelln[m]{V}{f(\bs)}} }
        \leq
        (m+1)^{\cardinality{\cX}\cardinality{\cY}}
        \exp\left\{ -m \CondEntropy{V}{\Pch} \right\}
    \end{align*}

    \paragraph{Bounding the sum of numerator}
    Note that since $\bs \in G(Q, \Pch)$, 
    \begin{equation}
        \by \in 
            \TypeShell{V}{f(\bs)} 
            \cap 
            \hB(\bs, D) 
        \Rightarrow
        \bs \in B(g_{J;n}(\by), Q, D)
        \cap 
        G(Q, \Pch),
        \label{eq:equivalence}
    \end{equation}
    hence any $\by$ will be counted at most $\cardinality{B(g_{J;n}(\by), Q, D) \cap G(Q, \Pch)}$ 
    times. According to \Cref{lemma:d-ball-size}, this is upper bounded by 
    $
        B_u 
        = 
        (n+1)^{\cardinality{\cS}{\cardinality{\hat{\cS}}}}
            \exp \left\{ n \left[ H(Q) - R(Q, D) \right] \right\}.
    $
    In addition, it is obvious that 
    $$
    \bigcup_{\bs_i \in G(Q, \Pch)} \TypeShell{V}{f(\bs_i)} \cap \hB(\bs_i, D)
    \subset
    \TypeSetn{\Psi},
    $$
    where 
    $\Psi = {\Pch}{V}$ is the channel output distribution corresponding to $\Pch$.
    Therefore,
    \begin{align*}
        \frac{1}{\cardinality{G(Q, \Pch)}}
        \sum_{\bs_i \in G(Q, \Pch)}
            \cardinality{ 
                \TypeShell{V}{f(\bs_i)} 
                \cap 
                \hB(\bs_i, D)
                }
        &\leq
        \frac{(n+1)^{\cardinality{\cX}+1}}{\cardinality{\TypeSetn{Q}}}
        B_u
        \cardinality{ 
            \bigcup_{\bs_i \in \TypeSetn{Q}}
                \TypeShell{V}{f(\bs_i)} 
                \cap 
                \hB(\bs_i, D)
                }
        \\
        &\leq
        \frac{(n+1)^{\cardinality{\cX}+1}}{\cardinality{\TypeSetn{Q}}}
        B_u
        \cardinality{ \TypeSetn{\Psi} }.
    \end{align*}
    Noting
    \begin{alignat*}{3}
        (n+1)^{-\cardinality{\cS}}
        \exp\left\{ n \Entropy{Q} \right\}
        &\leq
        \cardinality{\TypeSetn{Q}}
        %&\leq
        %\exp\left\{ n \Entropy{Q} \right\}
        \\
        %(m+1)^{-\cardinality{\cX}}
        %\exp\left\{ m \Entropy{\Psi} \right\}
        %&\leq
        \cardinality{\TypeSetn[m]{\Psi}}
        &\leq
        \exp\left\{ m \Entropy{\Psi} \right\},
    \end{alignat*}
    we have
    \begin{align*}
        \navelog 
        \left[ 
        \frac{(n+1)^{\cardinality{\cX}+1}}{\cardinality{\TypeSetn{Q}}}
        B_u
        \cardinality{ \TypeSetn{\Psi} }
        \right]
        &\leq
        \frac{\cardinality{\cX}+1}{n} \log (n+1)
        +
        \frac{\cardinality{\cS}}{n} \log (n+1)
        - \Entropy{Q} 
        \\
        &\quad
        + \frac{\cardinality{\cS}\cardinality{\hat{\cS}}}{n} \log (n+1)
        + H(Q)
        - R(Q, D)
        \\
        &\quad
        + \rho H(\Psi)
        \\
        &
        \leq 
        \rho H(\Psi) - R(Q,D)
        \\
        &\quad
        + \frac{\cardinality{\cS}\cardinality{\hat{\cS}}}{n} \log (n+1)
        + \frac{\cardinality{\cX}+1}{n} \log (n+1)
        + \frac{\cardinality{\cS}}{n} \log (n+1)
    \end{align*}
    Combining the bounds for both numerator and denominator, we have
    \begin{align*}
        &\navelog
        \left[ 
        \frac{1}{\cardinality{G(Q, \Pch)}}
        \sum_{\bs_i \in G(Q, \Pch)}
        \frac{ 
            \cardinality{ 
                \TypeShell{V}{f(\bs_i)} 
                \cap 
                \hB(\bs_i, D)
                }
        }
        {
            \TypeShell{V}{f(\bs_i)}
        }
        \right]
        \\
        \leq&\,
        \rho H(\Psi) - \rho \CondEntropy{V}{\Pch}
        - R(Q,D) 
        \\
        &
        + \frac{\cardinality{\cX}\cardinality{\cY}}{m} \log (m+1)
        + \frac{\cardinality{\cS}\cardinality{\hat{\cS}}}{n} \log (n+1)
        + \frac{\cardinality{\cX}+1}{n} \log (n+1)
        + \frac{\cardinality{\cS}}{n} \log (n+1)
    \end{align*}
    Note $m = \floor{\rho n} \leq \rho n$, let
    \begin{align}
        p(n) 
        = 
        (\rho n+1)^{\rho n \cardinality{\cX}\cardinality{\cY}}
        (n+1)^{n \left[ 
        (\cardinality{\cS}\cardinality{\hat{\cS}})
        (\cardinality{\cX}+1)
        (\cardinality{\cS})
        \right]},
        \label{eq:poly_term}
    \end{align}
    and the proof is completed. 
\end{IEEEproof}

\section{Continuity of the mutual information function}
In this section we show the continuity of the mutual information function, which shows that for
investigation in dispersion, arguments based on types is essentially the same as arguments based on
general probability distributions.

\begin{lemma}     
    \label{lemma:mutual_info_cont}
    For $P, Q\in\PDSpaceX$, if $\infnorm{P-Q}\leq \delta \leq 1/(2\cardinality{\cX}\cardinality{\cY})$, then
    \begin{equation*}
        \abs{\MIPW{P}{W} - \MIPW{Q}{W}}
        \leq
        \delta \cardinality{\cX}\log \cardinality{\cY}
        -\cardinality{\cY}\cardinality{\cX}\delta \log \cardinality{\cX}\delta.
    \end{equation*}
    Therefore, when $\delta = \BigTheta{\frac{1}{n}}$, 
    \[
    \abs{\MIPW{P}{W} - \MIPW{Q}{W}} = \BigO{\frac{\log n}{n}}
    \]
\end{lemma}
\begin{IEEEproof}
    Let $P_Y = \PDProd{P}{W}_Y$ and $Q_Y = \PDProd{Q}{W}_Y$, note
    \begin{align*}
        \onenorm{P_Y - Q_Y}
        &\leq
        \delta \cardinality{\cX}\cardinality{\cY}.
    \end{align*}
    Let $\delta' = \cardinality{\cX}\delta$, then Lemma 1.2.7 in~\cite{CsiszarBook} shows,
    \begin{align*}
        \abs{\MIPW{P}{W} - \MIPW{Q}{W}}
        &=
        \abs{
        \left(  
            \Entropy{P_Y} - \CondEntropy{W}{P}
        \right)
        -
        \left(  
        \Entropy{Q_Y} - \CondEntropy{W}{Q}
        \right)
        }
        \\
        &\leq
        \abs{ \left(  
            \Entropy{P_Y} - \Entropy{Q_Y} 
        \right) }
        +
        \abs{ \left(  
            \CondEntropy{W}{P} - \CondEntropy{W}{Q}
        \right) }
        \\
        &\leq 
        -\cardinality{\cY}\delta' \log \delta'
        +
        \delta \cardinality{\cX}\log \cardinality{\cY}
        \\
        &=
        \delta \cardinality{\cX}\log \cardinality{\cY}
        -\cardinality{\cY}\cardinality{\cX}\delta \log \cardinality{\cX}\delta.
    \end{align*}
\end{IEEEproof}

\section{Elementary Probability Inequalities}
In this section we prove several simple probability inequalities used in our derivation.

\begin{lemma}
Let $A$ and $B$ be two (generally dependent) random variables and let $c$ be a constant. Then for any 
values $\Gamma_1,\Gamma_2,\Gamma_3,\Gamma_4$,
the following holds:
\begin{align}
    \Prob{A+B > c} &\leq \Prob{A > c-\Gamma_1} + \Prob{B > \Gamma_1},
    \label{eq:prob_geq_upper}
    \\
    \Prob{A+B > c} &\geq \Prob{A > c+\Gamma_2} - \Prob{B < -\Gamma_2},
    \label{eq:prob_geq_lower}
    \\
    \Prob{A+B < c} &\leq \Prob{A < c+\Gamma_3} + \Prob{B < -\Gamma_3},
    \label{eq:prob_leq_upper}
    \\
    \Prob{A+B < c} &\geq \Prob{A < c-\Gamma_4} - \Prob{B > \Gamma_4}.
    \label{eq:prob_leq_lower}
\end{align}
\end{lemma} 

\begin{proof}
To show \Cref{eq:prob_geq_upper}, let
$\cE_{A} = \Set{A > c - \Gamma_1}$,
$\cE_{B} = \Set{B > \Gamma_1}$,
and 
$\cE = \Set{A + B > c}$.
Note that
\[ 
{\cE_{A}}^c \bigcap {\cE_{B}}^c
\subseteq
\cE^c,
\]
hence by De Morgan's law,
\[
{\cE_{A}} \bigcup {\cE_{B}}
\supseteq
\cE.
\]
We prove \Cref{eq:prob_geq_upper}
by the union bound
\[
\Prob{ \cE } 
\leq 
\Prob{\cE_{A}} + \Prob{\cE_{B}}.
\]

Apply
\Cref{eq:prob_geq_upper} on $-A, -B, -c$ and $\Gamma_2$, we obtain 
\Cref{eq:prob_geq_lower} after rearrangement.

Subtract 1 from both sides of \Cref{eq:prob_geq_lower} and replace $\Gamma_2$ by $\Gamma_3$, we
obtain \Cref{eq:prob_leq_upper} after rearrangement.

Apply
\Cref{eq:prob_geq_lower} on $-A, -B, -c$ and $\Gamma_4$, we obtain 
obtain \Cref{eq:prob_leq_lower} after rearrangement.

\end{proof}

\bibliographystyle{unsrt}
\bibliography{mybib}

\end{document}